\documentclass[aps,pra,showpacs]{revtex4}
\usepackage{graphicx}
\begin{document}

\title{Microscopic toy model for Cavity dynamical Casimir effect}
\author{I M de Sousa$^1$ and A V Dodonov$^{1,2}$}

\affiliation{$^1$ Institute of Physics,
University of Brasilia, 70910-900, Brasilia, Federal District, Brazil}

\affiliation{$^2$ International Center for Condensed Matter Physics,
University of Brasilia, 70910-900, Brasilia, Federal District, Brazil}

\begin{abstract}
We develop a microscopic toy model for Cavity dynamical Casimir effect
(DCE), namely, the photon generation from vacuum due to a nonstationary
dielectric slab in a fixed single mode cavity. We represent the slab by $%
N\gg 1$ noninteracting two-level atoms coupled to the field via the standard
dipole interaction. We show that the DCE is contained implicitly in the
light--matter interaction Hamiltonian when its parameters are externally
prescribed functions of time. We also predict several new phenomena, such as
saturation of the photon growth due to effective Kerr nonlinearity,
generation of pairs of atomic excitations instead of photons
(\textquotedblleft Inverse DCE\textquotedblright) and coherent annihilation
of pair of system excitations due to the atomic modulation
(\textquotedblleft Anti-DCE\textquotedblright). These results are extended
to the circuit QED architecture, where similar effects can be implemented
with a single qubit providing an alternative way to generate cavity and atom--field
entangled states.
\end{abstract}

\maketitle

\section{Introduction}

The term \textquotedblleft dynamical Casimir effect\textquotedblright\ (DCE)
is used nowadays for a rather wide group of phenomena characterized by
creation of quanta from the initial vacuum state of some field due to
time-dependent variations of the geometry or material properties of a
macroscopic or mesoscopic system (see \cite%
{book,vdodonov,revDal,nori,CAMOP-me} for recent reviews). In particular,
Cavity DCE \cite{2} denotes the process of photon generation from the
electromagnetic vacuum (and other initial states) in cavities due to the
motion of some wall or the time-modulation of the material properties (e.g.,
dielectric permittivity or conductivity) of the wall or a medium inside the
cavity \cite{bound2,diss4}. An analog of Cavity DCE was recently implemented
experimentally in the solid state architecture known as circuit Quantum
Electrodynamics (circuit QED \cite{cir1,cir2,cir3}), where a Josephson
metamaterial was embedded in a low-Q microwave cavity, permitting the
modulation of the cavity effective length via external magnetic field \cite%
{meta}.

Although Cavity DCE has been studied theoretically for more than four
decades, some aspects of this phenomenon are still not completely clear. A
particular issue we approach here is the asymptotic behavior of photon
generation: while some models predict the saturation of the intra-cavity
photon number \cite{sat1,sat2}, other predict exponential photon growth even
in the presence of moderate dissipation \cite%
{diss1,diss2,diss3,diss4,diss5,diss6}. This controversy can be resolved by
constructing a full microscopic model for the interaction between the
quantized electromagnetic field and moving or time-modulated objects
constituted of individual atoms. Some steps along this line were taken in
\cite{asym1,asym2}, yet the majority of studies employs time-varying
boundary conditions for the cavity field to bypass the complicated
light--matter interaction at the interface \cite%
{book,nori,bound1,bound2,bound3,bound4,bound5,bound6,bound7}.

In this paper we utilize the general mathematical description of
nonstationary circuit QED systems formulated recently in \cite{JPA} to
develop a microscopic toy model for Cavity DCE. Our study is motivated by
the following intuition: since the boundary conditions are just a
mathematical artifact to manage the interaction between photons and a large
number of atoms, DCE should ultimately originate from the basic form of
light--matter interaction with nonstationary parameters. So we consider the
special case of Cavity DCE implemented with a dielectric slab having
externally prescribed motion and material properties. The slab is portrayed
as an ensemble of $N$ two-level atoms with unspecified transition
frequencies and coupling strengths that interact with the field via the
standard dipole Hamiltonian with time-dependent parameters \cite{schleich}.
We use the time-independent boundary conditions to quantize the cavity field
in a standard manner, while the interaction between the arbitrarily
modulated atoms and photons is treated microscopically.

After cumbersome calculations we arrive at simple mathematical expressions
that generalize the common DCE description in single-mode cavities \cite%
{bound2,zeilinger}. In particular, we express the photon generation rate in
terms of the microscopic parameters, show that the photon growth and amount
of squeezing are limited due to effective Kerr nonlinearity and point out
that Cavity DCE occurs even for a single atom. Moreover, we discuss how
external classical pumping can significantly enhance the photon generation
from vacuum for suitable choices of the pump phase \cite{seed}. Since our
model is quite general, we also apply it to situations where all the system
parameters are known, such as a cloud of cold polar molecules \cite%
{polar1,polar2} or superconducting qubits \cite{cir3,cir4,cir5}. New effects
arising from periodic external modulations are analyzed: generation of pairs
of atomic excitations from vacuum (\textquotedblleft Inverse
DCE\textquotedblright ), coherent annihilation of a pair of system
excitations (\textquotedblleft Anti-DCE\textquotedblright ) and generation
of entangled light--matter states.

This paper is organized as follows. In section \ref{formulation} we
formulate our problem and in section \ref{toy} we develop the toy model for
cavity DCE, presenting the analytical and numerical results. In section \ref%
{clouds} we extend our analysis to cold atomic clouds, where all the atomic
parameters are known and, in principle, can be modulated externally. In
section \ref{qubit} we repeat this analysis for the case of a single
two-level atom, discussing the Anti-DCE behavior and studying some
applications in the area of circuit QED. Section \ref{conclusions} contains
the conclusions. This paper contains two extensive appendices: in \ref{an}
we give the thorough analytical description for the case $N\gg 1$ in the
Heisenberg picture, while in \ref{circuitqed} we do the same for $N=1$ in
the Schr\"{o}dinger picture.

\section{Mathematical formulation of the problem}

\label{formulation}

We quantize the cavity field using the standard methods with
time-independent boundary conditions \cite{schleich,vogel}. The annihilation
and creation operators $\hat{a}$ and $\hat{a}^{\dagger }$ do not depend
explicitly on time, so the vacuum state defined as $\hat{a}|0\rangle =0$
\cite{book,nori} is the same for all times, unlike the case of a cavity with
moving walls for which the field state depends on the instantaneous
frequency \cite{diss2}. We consider a small dielectric slab located at an
arbitrary position within the cavity, as depicted in figure 1. The
dielectric slab is subject to pre-determined motion with small amplitude,
and its material properties (e.g., dielectric permittivity) can be adjusted
externally by some bias (represented by the laser beam in the figure). From
the microscopic point of view, this problem corresponds to a predetermined
motion of an atomic cloud whose internal properties are modulated
externally. For consistency, the generation of photons from vacuum in this
particular example of DCE should be contained intrinsically within any
formulation of the light--matter interaction.

\begin{figure}[tbh]
\begin{center}
\includegraphics[natwidth=0.5\textwidth,natheight=0.3\textwidth]{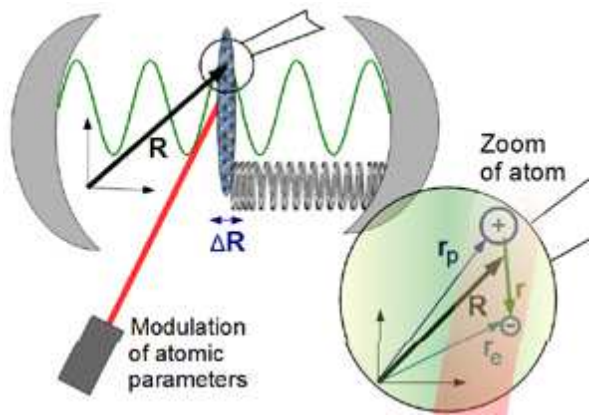} {}
\end{center}
\caption{Artistic view of DCE due to a nonstationary
dielectric slab in a fixed single-mode cavity. The dielectric slab (pictured
as a set of $N$ noninteracting Hydrogen atoms) oscillates according to an
external law of motion, while its dielectric properties can be modulated
externally via electric or magnetic fields. The harmonic wave represents the
time-independent cavity mode function; red beam represents the modulation of
the material properties of the dielectric. The zoom shows an individual atom
containing one proton and one electron, whose center-of-mass coordinate $%
\mathbf{R}$ changes due to the prescribed motion. }
\label{f1}
\end{figure}

We consider the simplest microscopic model for the dielectric slab -- a set
of $N$ non-interacting Hydrogen atoms, as shown in the zoom of figure 1.
First we recapitulate the interaction of a single atom with the field. Each
atom consists of a proton (electron), described by the position operator $%
\mathbf{\hat{r}}_{p}$ ($\mathbf{\hat{r}}_{e}$), with mass $m_{p}$ ($m_{e}$)
and charge $e$ ($-e$). Introducing the center-of-mass (CM) position operator
$\mathbf{\hat{R}}=(m_{e}\mathbf{\hat{r}}_{e}+m_{p}\mathbf{\hat{r}}_{p})/M$,
where $M=m_{e}+m_{p}$ is the total atomic mass, we define the momentum
operator $\mathbf{\hat{P}=\hat{p}}_{e}+\mathbf{\hat{p}}_{p}$ associated with
the CM motion, where $\mathbf{\hat{p}}_{p}$ ($\mathbf{\hat{p}}_{e}$) is the
canonical momentum operator of the proton (electron). Furthermore, one
introduces the relative coordinate between the proton and electron $\mathbf{%
\hat{r}}=\mathbf{\hat{r}}_{e}-\mathbf{\hat{r}}_{p}$ and the momentum $%
\mathbf{\hat{p}}=(m_{p}\mathbf{\hat{p}}_{e}+m_{e}\mathbf{\hat{p}}_{p})/M$
associated with the relative motion of the reduced mass $\mu =m_{e}m_{p}/M$.
As a result, we can decompose the dynamics into the motion of CM and the
relative motion, with the total kinetic energy given by $\mathbf{\hat{p}}%
_{e}^{2}/(2m_{e})+\mathbf{\hat{p}}_{p}^{2}/(2m_{p})=\mathbf{\hat{P}}%
^{2}/(2M)+\mathbf{\hat{p}}^{2}/(2\mu )$.

We treat the light--matter interaction in the first-order dipole
approximation, assuming that the dimensions of the atom are much smaller
than the wavelength of the cavity mode. The minimal coupling Hamiltonian in
the Coulomb gauge is minutely deduced in \cite{schleich}. Considering that
the CM motion is prescribed externally, with $\mathbf{P}$ and $\mathbf{R}$
given by known functions of time, it reads
\begin{eqnarray}
\hat{H}^{(1)} &=&\frac{\mathbf{P}^{2}}{2M}+\hat{H}_{f}+\hat{H}_{a}-e%
\mathbf{\hat{r}}\cdot \left[ \mathbf{\hat{E}}+\mathbf{\dot{R}}\times \mathbf{%
\hat{B}}\right] +e^{2}\left( \frac{1}{8\mu }-\frac{1}{M}\right) \left[
\mathbf{\hat{r}}^{2}\mathbf{\hat{B}}^{2}-(\mathbf{\hat{r}}\cdot \mathbf{\hat{%
B})}^{2}\right]  \nonumber \\
&&-\frac{e(m_{p}-m_{e})}{2M}\left[ \frac{1}{\mu }\,\mathbf{\hat{B}}\cdot (%
\mathbf{\hat{r}}\times \mathbf{\hat{p})}+(\mathbf{\hat{r}}\cdot \mathbf{%
\nabla }_{\mathbf{R}})\mathbf{\hat{r}}\cdot \mathbf{\hat{E}}\right] ~.
\label{H}
\end{eqnarray}%
Here $\hat{H}_{f}=\hbar \omega \hat{n}$ is the cavity free Hamiltonian,
where $\omega $ is the frequency and $\hat{n}=\hat{a}^{\dagger }\hat{a}$ is
the photon number operator. $\hat{H}_{a}=\mathbf{\hat{p}}^{2}/(2\mu )+V(%
\mathbf{\hat{r})}$ is the Hamiltonian of the atomic internal dynamics, where
$V(\mathbf{\hat{r})}=-e^{2}/(4\pi \varepsilon _{0}|\mathbf{\hat{r}|})$ is
the Coulomb interaction energy and $\varepsilon _{0}$ is the permittivity of
vacuum. The electric and magnetic intracavity fields are%
\begin{equation}
\mathbf{\hat{E}}\left( \mathbf{R}\right) =i\sqrt{\frac{\hbar \omega }{%
2\varepsilon _{0}V}}\mathbf{u}\left( \mathbf{R}\right) \left( \hat{a}-\hat{a}%
^{\dagger }\right) ~
\end{equation}%
\begin{equation}
\mathbf{\hat{B}}\left( \mathbf{R}\right) =\sqrt{\frac{\hbar }{2\varepsilon
_{0}V\omega }}\left( \mathbf{\nabla }\times \mathbf{u}\left( \mathbf{R}%
\right) \right) \left( \hat{a}+\hat{a}^{\dagger }\right) \,,
\end{equation}%
where $V$ is the mode volume and $\mathbf{u}\left( \mathbf{R}\right) $ is
the dimensionless mode function determined from the time-independent
boundary conditions on the walls. In the stationary case, when $\mathbf{\dot{%
R}=0}$, one usually neglects the contributions containing the magnetic field
and the gradient of the electric field in Hamiltonian (\ref{H}), recovering
the standard dipole interaction term $-e\mathbf{\hat{r}}\cdot \mathbf{\hat{E}%
}$. However, in the nonstationary regime $\mathbf{\dot{R}\neq 0}$ all the
terms must be taken into account \cite{schleich}.

For our toy model we take into consideration only the two atomic levels
near-resonant with the cavity frequency, restricting the atomic dynamics to
the \textquotedblleft ground\textquotedblright\ and \textquotedblleft
excited\textquotedblright\ states $|g\rangle $ and $|e\rangle $,
respectively. Hence the atomic Hamiltonian reads $\hat{H}_{a}=\hbar \Omega
\hat{\sigma}_{z}/2$, where $\Omega $ is the transition frequency and $\hat{%
\sigma}_{z}=|e\rangle \langle e|-|g\rangle \langle g|$ is the Pauli
operator. For the two-level approximation to hold we must have $|\omega
-\Omega |\ll \omega $. The position operator can be written as $\mathbf{\hat{%
r}=r}_{0}\hat{\sigma}_{+}+\mathbf{r}_{0}^{\ast }\hat{\sigma}_{-}$, where $%
\mathbf{r}_{0}=\langle e|\mathbf{\hat{r}}|g\rangle $ is the off-diagonal
matrix element and the Pauli ladder operators are $\hat{\sigma}%
_{+}=|e\rangle \langle g|$ and $\hat{\sigma}_{-}=|g\rangle \langle e|$. In
this case $\mathbf{\hat{r}}^{2}=|\mathbf{r}_{0}|^{2}$ and the square of the
magnetic field operator $\propto (\hat{a}+\hat{a}^{\dagger })^{2}$ appears
naturally in Hamiltonian (\ref{H}).

Hence the simplest model for a nonstationary dielectric slab in a stationary
cavity is described by the general Hamiltonian of the form (we set $\hbar =1$%
)%
\begin{equation}
\hat{H}=\omega \hat{n}+\sum_{l=1}^{N}\left[ \frac{\Omega }{2}\hat{\sigma}%
_{z}^{(l)}+g(\hat{a}+\hat{a}^{\dagger })(\hat{\sigma}_{+}^{(l)}+\hat{\sigma}%
_{-}^{(l)})\right] +i\chi (\hat{a}^{\dagger 2}-\hat{a}^{2})+id(\hat{a}%
^{\dagger }-\hat{a})~,  \label{H1}
\end{equation}%
where the index $l$ labels the identical noninteracting atoms. The
renormalized cavity frequency $\omega $ is constant, while the atomic
transition frequency $\Omega $, the atom--cavity coupling strength$~g$ and
the \textquotedblleft squeezing coefficient\textquotedblright\ $\chi $ are
regarded as externally prescribed functions of time. This occurs both due to
the motion of the slab and the external \emph{in situ} modulation of the
atomic properties, though here we do not pursue the exact dependence. The
last term on the right-hand side (RHS) of equation (\ref{H1}) accounts for
the classical one-photon pumping of the cavity field \cite{cir1}, included
for generality and to study how DCE can be enhanced by an additional
coherent drive.

To understand the emergence of DCE from the microscopic viewpoint we do not
need to know the exact relation between the parameters of Hamiltonians (\ref%
{H}) and (\ref{H1}), since for a weak external perturbation of the system we
can write
\begin{equation}
X=X_{0}+\varepsilon _{X}\sum_{j}w_{X}^{(j)}\sin \left( \eta ^{(j)}t+\varphi
_{X}^{(j)}\right) ~,~X=\{\omega ,\Omega ,g,\chi ,d\}\,,  \label{jones}
\end{equation}%
where $X_{0}$ is the bare value and $\varepsilon _{X}\geq 0$ is the
modulation depth of $X$. The sum runs over all the present modulation
frequencies $\eta ^{(j)}$; we can write it as $\sum_{j}=\sum_{j}^{\prime
}+\sum_{j}^{\prime \prime }$, where $\sum_{j}^{\prime }$ denotes the sum
over \textquotedblleft fast\textquotedblright\ modulation frequencies, $\eta
^{(j\prime )}\gtrsim \omega _{0}$, and $\sum_{j}^{\prime \prime }$ -- over
\textquotedblleft slow\textquotedblright\ modulation frequencies, $\eta
^{(j\prime \prime )}\ll \omega _{0}$. Parameters $w_{X}^{(j)}\geq 0$ and $%
\varphi _{X}^{(j)}$ are the relative weights and phase constants
corresponding to the modulation of $X$ at frequency $\eta ^{(j)}$. For the
classical pump we set $d_{0}=0$, and we included the modulation of $\omega $
in equation (\ref{jones}) for the sake of generality. For the future use we
define the complex modulation depth $\varepsilon _{X}^{(j)}$ that includes
the weight and the phase of $X$-modulation at frequency $\eta ^{(j)}$%
\begin{equation}
\varepsilon _{X}^{(j)}\equiv \varepsilon _{X}w_{X}^{(j)}\exp [i\phi
_{X}^{(j)}]~,~X=\{\omega ,\Omega ,g,\chi ,d\}~.
\end{equation}%
Throughout the paper the notation $\varepsilon _{X}^{(j\prime )}$ and $%
\varepsilon _{X}^{(j\prime \prime )}$ stands for the complex modulation
depths corresponding to fast and slow modulation frequencies, respectively.

\section{Toy model for DCE with a dielectric slab}

\label{toy}

As a toy model for DCE we consider a fixed cavity of known frequency $\omega
_{0}$ that contains $N$ identical two-level atoms. The atomic transition
frequency $\Omega $ and the coupling strength $g$ are unknown, but in order
to represent the dielectric slab the difference $\omega _{0}-\Omega $ must
be large compared to the coupling strength, $|\omega _{0}-\Omega |\gg |g|$.
Due to the external perturbation the parameters $\Omega $, $g$ and $\chi $
vary according to equation (\ref{jones}), and we consider the general case
when the three parameters can change simultaneously. For the macroscopic
slab we consider $N\gg 1$ and define the collective operators via the
Holstein--Primakoff transformation \cite{HP}%
\begin{equation}
\sum_{l=1}^{N}\hat{\sigma}_{+}^{(l)}=\hat{b}^{\dagger }(N-\hat{b}^{\dagger }%
\hat{b})^{1/2}~~,~\sum_{l=1}^{N}\hat{\sigma}_{-}^{(l)}=(N-\hat{b}^{\dagger }%
\hat{b})^{1/2}\hat{b}~,~\sum_{l=1}^{N}\hat{\sigma}_{z}^{(l)}=2\hat{b}%
^{\dagger }\hat{b}-N~,
\end{equation}%
where the ladder operators $\hat{b}$ and $\hat{b}^{\dagger }$ satisfy the
bosonic commutation relation $[\hat{b},\hat{b}^{\dagger }]=1$. To the first
order in $\hat{b}^{\dagger }\hat{b}/N$ the Hamiltonian for our toy model
reads%
\begin{equation}
\hat{H}=\omega _{0}\hat{n}+\Omega \hat{b}^{\dagger }\hat{b}+\tilde{g}(\hat{a}%
+\hat{a}^{\dagger })(\hat{b}+\hat{b}^{\dagger })+i\chi (\hat{a}^{\dagger 2}-%
\hat{a}^{2})-\frac{\tilde{g}}{2N}(\hat{a}+\hat{a}^{\dagger })(\hat{b}%
^{\dagger 2}\hat{b}+\hat{b}^{\dagger }\hat{b}^{2})~,  \label{NHN}
\end{equation}%
where we defined the collective coupling constant $\tilde{g}\equiv \sqrt{N}g$%
, so that $\tilde{g}_{0}=\sqrt{N}g_{0}$ and $\tilde{\varepsilon}_{g}=\sqrt{N}%
\varepsilon _{g}$ (we consider $g_{0}\geq 0$ without loss of generality). In
this paper the tilde over a c-number corresponds to the collective $N$-atoms
parameter. The Hamiltonian (\ref{NHN}) holds provided the inequality $%
\langle \hat{b}^{\dagger }\hat{b}\rangle \ll N$ is satisfied.

In the dispersive regime, $|\Delta _{-}|/2\gg \tilde{g}_{0}$, where $\Delta
_{-}=\omega _{0}-\Omega _{0}$ is the bare atom--field detuning, the
approximate solution in the Heisenberg picture is deduced in \ref{dispersive}%
:%
\begin{eqnarray}
\hat{a} &\simeq &e^{-i(\omega _{0}-\tilde{\delta}_{+}-\delta _{\chi })t}%
\left[ \hat{A}-i\hat{B}\frac{2\tilde{g}_{0}}{\Delta _{-}}e^{i(\Delta
_{-}/2-\delta _{\chi })t}\sin \left( \frac{\Delta _{-}t}{2}\right) \right]
\label{aa} \\
\hat{b} &\simeq &e^{-i(\Omega _{0}-\tilde{\delta}_{+})t}\left[ \hat{B}-i\hat{%
A}\frac{2\tilde{g}_{0}}{\Delta _{-}}e^{-i(\Delta _{-}/2-\delta _{\chi
})t}\sin \left( \frac{\Delta _{-}t}{2}\right) \right]  \label{bb}
\end{eqnarray}%
\begin{equation}
\Delta _{+}=\omega _{0}+\Omega _{0}~,~\tilde{\delta}_{\pm }=\frac{\tilde{g}%
_{0}^{2}}{\omega _{0}\pm \Omega _{0}}~,~\delta _{\chi }=\frac{4\chi _{0}^{2}%
}{\omega _{0}+\Omega _{0}}~.
\end{equation}%
$\hat{A}$ and $\hat{B}$ are independent bosonic ladder operators that obey
the Heisenberg equation of motion $id\hat{O}/dt=[\hat{O},\hat{H}_{eff}]$ ($%
\hat{O}=\hat{A},\hat{B}$) with the effective Hamiltonian%
\begin{equation}
\hat{H}_{eff}=\hat{H}_{G}+\hat{H}_{NG}~.  \label{Heff}
\end{equation}%
Here $\hat{H}_{G}$ contains the Gaussian part (quadratic terms in the
operators $\hat{A}$ and $\hat{B}$) and $\hat{H}_{NG}$ contains the
non-Gaussian part (quartic terms).

For the modulation frequency%
\begin{equation}
\eta ^{(D)}=2(\omega _{0}+\tilde{\delta}_{-}-\tilde{\delta}_{+}-\delta
_{\chi }-\zeta )~,  \label{eta}
\end{equation}%
where we introduced the small adjustable \textquotedblleft resonance
shift\textquotedblright\ $\zeta $ in order to perform the fine tuning of the
modulation frequency, we find%
\begin{equation}
\hat{H}_{G}=qe^{-2it\zeta }\left[ \hat{A}^{2}+2e^{-it\delta _{\chi }}\frac{%
\tilde{g}_{0}}{\Delta _{-}}\hat{A}\hat{B}+e^{-2it\delta _{\chi }}\left(
\frac{\tilde{g}_{0}}{\Delta _{-}}\right) ^{2}\hat{B}^{2}\right] +h.c.
\label{vista}
\end{equation}%
\begin{equation}
q=\frac{i\tilde{\delta}_{-}\Omega _{0}}{\left( \omega _{0}+\Omega
_{0}\right) }\left[ \frac{\varepsilon _{\Omega }^{(D)}}{2\Omega _{0}}-\frac{%
\tilde{\varepsilon}_{g}^{(D)}}{\tilde{g}_{0}}\right] -\frac{\varepsilon
_{\chi }^{(D)}}{2}  \label{vi3}
\end{equation}%
\begin{eqnarray}
\hat{H}_{NG} &=&-\frac{\tilde{\delta}_{-}}{2N}\left[ \frac{3\tilde{g}%
_{0}^{2}}{\Delta _{-}^{2}}\hat{A}^{\dagger 2}\hat{A}^{2}-\hat{B}^{\dagger 2}%
\hat{B}^{2}+\frac{8\tilde{g}_{0}}{\Delta _{-}}\hat{A}\hat{B}^{\dagger 2}\hat{%
B}e^{i\delta _{\chi }t}+2\hat{A}^{\dagger }\hat{A}\hat{B}^{\dagger }\hat{B}%
\right.  \label{vista2} \\
&&\left. -\frac{6\tilde{g}_{0}^{2}}{\Delta _{-}^{2}}\hat{A}^{\dagger 2}\hat{B%
}^{2}e^{-2i\delta _{\chi }t}-\frac{4\tilde{g}_{0}}{\Delta _{-}}\hat{A}%
^{\dagger 2}\hat{A}\hat{B}e^{-i\delta _{\chi }t}\right] +i\frac{\tilde{\delta%
}_{-}}{2N}\frac{\tilde{\varepsilon}_{g}^{(D)}}{\tilde{g}_{0}}e^{-2i\zeta t}%
\left[ \hat{A}^{2}\hat{B}^{\dagger }\hat{B}\right.  \nonumber \\
&&+\frac{3\tilde{g}_{0}^{2}}{2\Delta _{-}^{2}}\hat{A}^{\dagger }\hat{A}^{3}+%
\frac{2\tilde{g}_{0}}{\Delta _{-}}\hat{A}\hat{B}^{\dagger }\hat{B}%
^{2}e^{-i\delta _{\chi }t}-\frac{\tilde{g}_{0}}{\Delta _{-}}\hat{A}^{\dagger
}\hat{A}^{2}\hat{B}e^{-i\delta _{\chi }t}-\frac{\tilde{g}_{0}}{\Delta _{-}}%
\hat{A}^{3}\hat{B}^{\dagger }e^{i\delta _{\chi }t}  \nonumber \\
&&-\frac{2\tilde{g}_{0}^{2}}{\Delta _{-}^{2}}\hat{A}^{\dagger }\hat{A}\hat{B}%
^{2}e^{-2i\delta _{\chi }t}\left. +\frac{\tilde{g}_{0}^{2}}{\Delta _{-}^{2}}%
\hat{B}^{\dagger }\hat{B}^{3}e^{-2i\delta _{\chi }t}-\frac{\tilde{g}_{0}^{3}%
}{\Delta _{-}^{3}}\hat{A}^{\dagger }\hat{B}^{3}e^{-3i\delta _{\chi }t}\right]
+h.c.  \nonumber
\end{eqnarray}

These results were obtained under a series of approximations. First, the
detuning and the modulation depth of the atomic transition frequency must be
small, $\varepsilon _{\Omega },|\Delta _{-}|\ll \omega _{0}$, while the
modulation depth of the atom-field coupling strength is $\tilde{\varepsilon}%
_{g}\lesssim \tilde{g}_{0}$. Second, there are some restraints on the number
of excitations in the atoms--field system for which our approach is accurate:%
\begin{equation}
\frac{\langle \hat{b}^{\dagger }\hat{b}\rangle }{N},\sqrt{\langle \hat{a}%
^{\dagger }\hat{a}\rangle +\langle \hat{b}^{\dagger }\hat{b}\rangle }\left\{
\frac{\tilde{g}_{0}}{\omega _{0}},\frac{\tilde{\varepsilon}_{g}}{\Delta _{-}}%
,\frac{|\chi _{0}|}{\omega _{0}},\frac{\varepsilon _{\chi }}{\omega _{0}},%
\frac{\tilde{g}_{0}\varepsilon _{\Omega }}{\Delta _{-}\omega _{0}},\frac{%
\tilde{g}_{0}\varepsilon _{\chi }}{\Delta _{-}^{2}}\right\} \ll 1  \label{w1}
\end{equation}%
\begin{equation}
\frac{\tilde{g}_{0}}{\Delta _{-}}\sqrt{\langle \hat{a}^{\dagger }\hat{a}%
\rangle +\langle \hat{b}^{\dagger }\hat{b}\rangle }\left\{ \frac{\langle
\hat{b}^{\dagger }\hat{b}\rangle }{N},\frac{\sqrt{\langle \hat{b}^{\dagger }%
\hat{b}\rangle \langle \hat{a}^{\dagger }\hat{a}\rangle }}{N},\frac{\tilde{g}%
_{0}}{|\Delta _{-}|}\frac{\langle \hat{a}^{\dagger }\hat{a}\rangle }{N}%
\right\} \ll 1~.  \label{w2}
\end{equation}%
Third, in equation (\ref{eta}) there are small \textquotedblleft Systematic
error frequency shifts\textquotedblright\ (SEFS) $\Delta \eta $ that were
neglected in order to keep the formulae concise. They are of the order%
\begin{eqnarray}
O(\Delta \eta ) &\sim &\left\{ \tilde{\delta}_{-}\left( \frac{\varepsilon
_{\Omega }}{\omega _{0}}\right) ^{2},\tilde{\delta}_{-}\left( \frac{\tilde{%
\varepsilon}_{g}}{\tilde{g}_{0}}\right) ^{2},\tilde{\delta}_{-}\left( \frac{%
\varepsilon _{\chi }}{\Delta _{-}}\right) ^{2},\right.  \nonumber \\
&&\left. \quad \tilde{\delta}_{+}\left( \frac{\varepsilon _{\Omega }}{\Delta
_{-}}\right) ^{2},\tilde{\delta}_{+}\left( \frac{\Delta _{-}}{\omega _{0}}%
\right) ^{2},\frac{\varepsilon _{\chi }^{2}}{\omega _{0}},\frac{\chi _{0}^{2}%
}{\omega _{0}}\left( \frac{\Delta _{-}}{\omega _{0}}\right) ^{2}\right\} .
\end{eqnarray}%
Hence in the actual implementation of DCE one has to find experimentally the
exact modulation frequency by scanning within the range $\Delta \eta $, so
in part we introduced the adjustable resonance shift $\zeta $ to achieve
this fine tuning.

One can simplify the Hamiltonian (\ref{Heff}) a little further. Neglecting
the non-Gaussian terms we have%
\begin{equation}
\frac{d}{dt}\hat{B}=e^{it\delta _{\chi }}\frac{\tilde{g}_{0}}{\Delta _{-}}%
\frac{d}{dt}\hat{A}~,
\end{equation}%
so for $|q|\gg |\delta _{\chi }|$ we can write%
\begin{equation}
\hat{B}(t)\simeq \hat{B}(0)+e^{it\delta _{\chi }}\frac{\tilde{g}_{0}}{\Delta
_{-}}\left[ \hat{A}(t)-\hat{A}(0)\right] ~.  \label{fried}
\end{equation}%
Assuming that the cavity and the atoms were initially in the ground states
and substituting equation (\ref{fried}) into (\ref{Heff}), to the lowest
order in $\tilde{g}_{0}/\Delta _{-}$ we obtain the Hamiltonian%
\begin{equation}
\hat{H}_{eff}\simeq \left( qe^{-2it\zeta }\hat{A}^{2}+h.c.\right) -N\alpha %
\left[ (\hat{A}^{\dagger }\hat{A})^{2}+\hat{A}^{\dagger }\hat{A}\right] ~,
\label{marty}
\end{equation}%
where%
\begin{equation}
\alpha =\frac{g_{0}^{4}}{\Delta _{-}^{3}}
\end{equation}%
is the effective Kerr nonlinearity strength due to a single two-level atom.
Defining the phase $\phi _{q}$ via the relation $q=i\left\vert q\right\vert
e^{i\phi _{q}}$ and introducing the new annihilation operator
\begin{equation}
\hat{a}_{r}=-ie^{i\phi _{q}/2}e^{-it\zeta }\hat{A}  \label{entre}
\end{equation}%
(that also satisfies $[\hat{a}_{r},\hat{a}_{r}^{\dagger }]=1$), the
evolution of $\hat{a}_{r}$ is governed by the time-independent\emph{\
Nonlinear DCE Hamiltonian}%
\begin{equation}
\hat{H}_{DCE}=\omega _{r}\hat{n}_{r}+\alpha _{r}\hat{n}_{r}^{2}+iq_{r}(\hat{a%
}_{r}^{\dagger 2}-\hat{a}_{r}^{2})~.  \label{seth0}
\end{equation}%
Here $\hat{n}_{r}=\hat{a}_{r}^{\dagger }\hat{a}_{r}$, $\omega _{r}=\left(
\zeta -N\alpha \right) $,$~\alpha _{r}=-N\alpha $ and $q_{r}=\left\vert
q\right\vert $. The term $iq_{r}(\hat{a}_{r}^{\dagger 2}-\hat{a}_{r}^{2})$,
which describes the simplest case of DCE in oscillating cavities \cite%
{bound2}, appears naturally in our derivation. The Hamiltonian (\ref{seth0})
is well known from Nonlinear Quantum Optics for describing (in the
interaction picture) a cavity that contains a Kerr medium and is
parametrically driven \cite%
{kerr1,milburn1,milburn2,kerr2,milburn3,milburn4,milburn5,kryuch2,kryuch1,leonski,kerr3,kryuch3}%
, so we call $\omega _{r}$ \textquotedblleft effective
detuning\textquotedblright .

Hence we were able to deduce microscopically the DCE from the most basic
form of light--matter interaction, equation (\ref{NHN}). It turns out that
DCE is described by the cumbersome non-Gaussian Hamiltonian given by
equations (\ref{vista}) -- (\ref{vista2}), and the standard expression for
cavity DCE is recovered only to the lowest order in $\tilde{g}_{0}/\Delta
_{-}$. Recalling that the auxiliary annihilation operators $\hat{A}$ and $%
\hat{B}$ are related to the physical annihilation operators $\hat{a}$ and $%
\hat{b}$ via relations (\ref{aa}) -- (\ref{bb}), we can formulate the first
new prediction of our toy model: the photon creation from vacuum is
accompanied by the excitation of the internal degrees of freedom of the
atoms in the slab, which becomes entangled with the cavity field. As stated
previously, we assume that the CM motion of the atoms is prescribed
externally, so our model does not contemplate the important back-action
effects of DCE on the motion of the slab \cite{revDal,back1,back2}.

The simplest \emph{realistic} description of Cavity DCE must include (at
least) the Kerr nonlinearity, as shown by equation (\ref{seth0}). Although
separately the DCE and Kerr Hamiltonians can be integrated in a
straightforward manner \cite{kerr1}, the general analytical solution for the
nonlinear DCE Hamiltonian is not known. To get qualitative insights about
the asymptotic dynamics of Hamiltonian (\ref{seth0}) we rewrite it in the
form of interaction picture parametric amplifier $\hat{H}_{DCE}=\hat{D}_{r}%
\hat{n}_{r}+iq_{r}(\hat{a}_{r}^{\dagger 2}-\hat{a}_{r}^{2})$, where the
overall detuning operator is $\hat{D}_{r}\equiv \omega _{r}+\alpha _{r}\hat{n%
}_{r}$. Treating the detuning as a c-number $\langle \hat{D}_{r}\rangle $,
the solution in the Heisenberg picture reads \cite{puri}%
\begin{equation}
\hat{a}_{r}\left( t\right) =\mathcal{F}^{\ast }\hat{a}_{r}\left( 0\right) +%
\mathcal{G}\hat{a}_{r}^{\dagger }\left( 0\right)
\end{equation}
\begin{equation}
\mathcal{F}\equiv \cosh \left( \mathcal{B}t\right) +i\frac{\langle \hat{D}%
_{r}\rangle }{\mathcal{B}}\sinh \left( \mathcal{B}t\right) ~,~\mathcal{G}%
\equiv 2\frac{q_{r}}{\mathcal{B}}\sinh \left( \mathcal{B}t\right) ~,~~%
\mathcal{B}=\sqrt{4q_{r}^{2}-\langle \hat{D}_{r}\rangle ^{2}}~.
\end{equation}%
So we arrive at the second new prediction of our model: asymptotic
exponential photon growth is impossible for nonzero $\alpha _{r}$, as for
any fixed value of $\omega _{r}$ the parameter $\mathcal{B}$ becomes
imaginary for $\langle \hat{n}_{r}\rangle \rightarrow \infty $. This results
solves the controversy about the long-time behavior of Cavity DCE,
supporting the finding \cite{sat1,sat2} that the photon generation is
limited even in the absence of dissipation.

To elucidate the system behavior for finite $\langle \hat{n}_{r}\rangle $ we
write the wavefunction corresponding to the Hamiltonian $\hat{H}_{DCE}$ as%
\begin{equation}
|\psi \rangle =\sum_{m=0}^{\infty }\exp [-it(\omega _{r}m+\alpha
_{r}m^{2})]c_{m}|m\rangle ~,
\end{equation}%
where $|m\rangle $ denotes the Fock state. The probability amplitudes obey
the differential equation%
\begin{equation}
\dot{c}_{m}=q_{r}[\sqrt{m(m-1)}e^{2it[\omega _{r}+2\alpha _{r}(m-1)]}c_{m-2}-%
\sqrt{(m+1)(m+2)}e^{-2it[\omega _{r}+2\alpha _{r}(m+1)]}c_{m+2}].
\label{seth}
\end{equation}%
One can easily solve the pair of equations connecting just the amplitudes $%
c_{K}\left( t\right) $ and $c_{K+2}\left( t\right) $ \cite{JPA}. For $%
c_{K+2}\left( 0\right) =0$ we get%
\begin{equation}
c_{K+2}(t)=e^{it\left[ \omega _{r}+2\alpha _{r}\left( K+1\right) \right] }%
\frac{q_{r}\sqrt{\left( K+1\right) \left( K+2\right) }}{R_{K}}\sin \left(
R_{K}t\right) c_{K}(0)~  \label{wr}
\end{equation}%
\begin{equation}
R_{K}=\sqrt{[\omega _{r}+2\alpha _{r}(K+1)]^{2}+q_{r}^{2}(K+1)(K+2)}~.
\end{equation}%
So the probability amplitude $c_{K+2}$ is approximately decoupled form $%
c_{K} $ when $q_{r}\sqrt{(K+1)(K+2)}\ll R_{K}$. For $2|\alpha _{r}|\left(
K+1\right) \gg |\omega _{r}|$ the decoupling condition becomes $q_{r}\ll
2|\alpha _{r}|$. Therefore, in order to generate many photons from vacuum
one must satisfy the condition $q_{r}\gg 2|\alpha _{r}|$. In the opposite
case, $q_{r}\lesssim |\alpha _{r}|$, we expect generation of a small number
of photons \cite{leonski}. For example, for $K=0$ and $q_{r}/|\alpha
_{r}|\ll 1$, the effective detuning must be adjusted to $\omega
_{r}=-2\alpha _{r}$ to optimize the coupling between the probability
amplitudes $c_{0}$ and $c_{2}$. As $q_{r}/|\alpha _{r}|$ increases one can
set $\omega _{r}=-2\left( K+1\right) \alpha _{r}$ to optimize the coupling
between the amplitudes $\{c_{K},c_{K+2}\}$ ($K=2,4,\ldots $), while the
off-resonant coupling between $\{c_{0},c_{2},\cdots ,c_{K}\}$ still allows a
substantial population of $c_{K}$ \cite{1}. So the question of utmost
practical interest is: what value of $\omega _{r}$, or equivalently, what
value of the adjustable resonance shift $\zeta $ optimizes the photon
generation from vacuum in the presence of Kerr nonlinearity? The answer will
given in the next subsection with the help of numerical simulations.

\subsection{Numerical results}

\begin{figure}[tbh]
\begin{center}
\includegraphics[width=.9\textwidth]{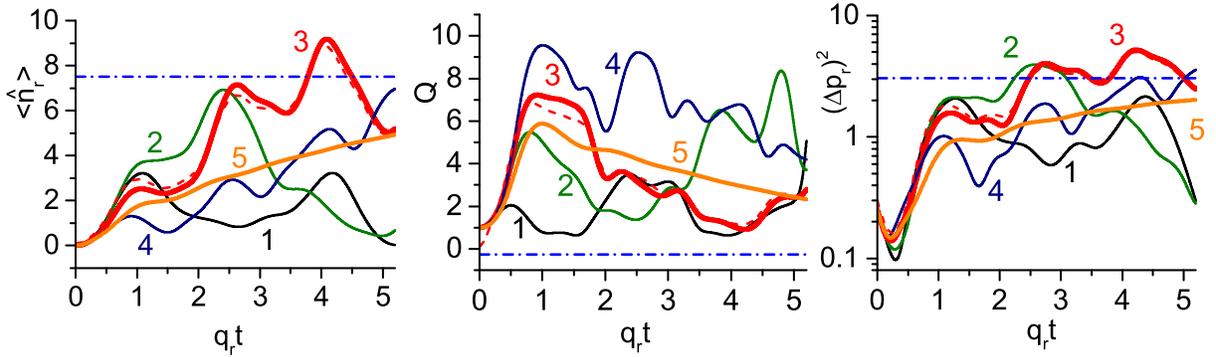} {}
\end{center}
\caption{Time behavior of the average photon number $%
\left\langle \hat{n}_{r}\right\rangle $, Mandel $Q$-factor and the variance
of the squeezed quadrature $\left( \Delta p_{r}\right) ^{2}$ obtained via
numerical integration of equation (\protect\ref{ssme1})\ for $q_{r}=3|%
\protect\alpha _{r}|$. For curves 1 -- 5 the initial state is the vacuum
state. For $\protect\kappa =0$ the curves are: $\protect\omega _{r}=0$ (1), $%
\protect\omega _{r}=-8\protect\alpha _{r}$ (2), $\protect\omega _{r}=-10%
\protect\alpha _{r}$ (3) and $\protect\omega _{r}=-12\protect\alpha _{r}$
(4). Line 5: $\protect\omega _{r}=-10\protect\alpha _{r}$ and $\protect%
\kappa =q_{r}$; the dot-dashed line indicates the asymptotic value in this
case. The dashed line corresponds to the initial thermal state with the
average photon number $\bar{n}=0.1$ and parameters $\protect\omega _{r}=-10%
\protect\alpha _{r}$, $\protect\kappa =0$. }
\label{fig2}
\end{figure}

We studied numerically how the Kerr nonlinearity affects the photon
generation from vacuum. For the sake of completeness we included the cavity
damping by means of the standard master equation at zero temperature \cite%
{schleich,vogel}%
\begin{equation}
\frac{d{\hat{\rho}}}{dt}=-i[\hat{H}_{DCE},\hat{\rho}]+\frac{\kappa }{2}%
\left( 2\hat{a}_{r}\hat{\rho}\hat{a}_{r}^{\dagger }-\hat{a}_{r}^{\dagger }%
\hat{a}_{r}\hat{\rho}-\hat{\rho}\hat{a}_{r}^{\dagger }\hat{a}_{r}\right) ,
\label{ssme1}
\end{equation}%
where $\hat{\rho}$ is the density operator, $\kappa $ is the cavity damping
rate and $\hat{H}_{DCE}$ is given by equation (\ref{seth0}). Strictly
speaking, the microscopic derivation of this master equation does not
contemplate the nonstationary case studied here, when the system parameters
vary rapidly with time and the counter-rotating terms play a fundamental
role \cite{bea}. Hence the solution of the master equation can only be used
to grasp qualitatively the overall effect of dissipation. The stationary
state of equation (\ref{ssme1}) can be calculated analytically using the
method of potential solutions for the corresponding Fokker-Planck equation
\cite{kryuch1,kryuch2}. However, as shown in figure \ref{fig2}, the
asymptotic solution is of little help for our problem because the cavity
field state during the time period of interest (initial times) may be very
different from the asymptotic one.

\begin{figure}[tbh]
\begin{center}
\includegraphics[width=0.6\textwidth]{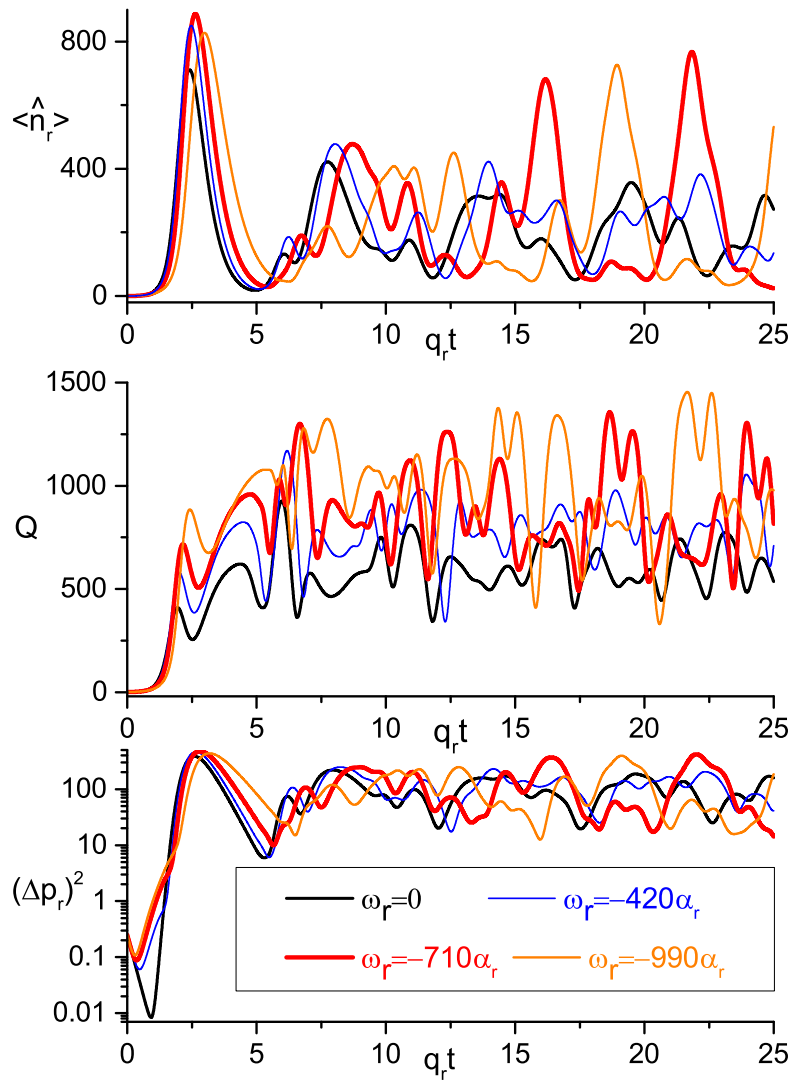} {}
\end{center}
\caption{Time behavior of $\left\langle \hat{n}%
_{r}\right\rangle $, $Q$ and $\left( \Delta p_{r}\right) ^{2}$ for the
initial vacuum state, $q_{r}=650|\protect\alpha _{r}|$, $\protect\kappa =0$
and different values of $\protect\omega _{r}$. Notice the irregular
collapse-revival behavior of $\left\langle \hat{n}_{r}\right\rangle $ and
the maximization of the average number of created photons for $\protect%
\omega _{r}^{(\max )}=-710\protect\alpha _{r}$.}
\label{fig3}
\end{figure}
\begin{figure}[tbh]
\begin{center}
\includegraphics[width=0.7\textwidth]{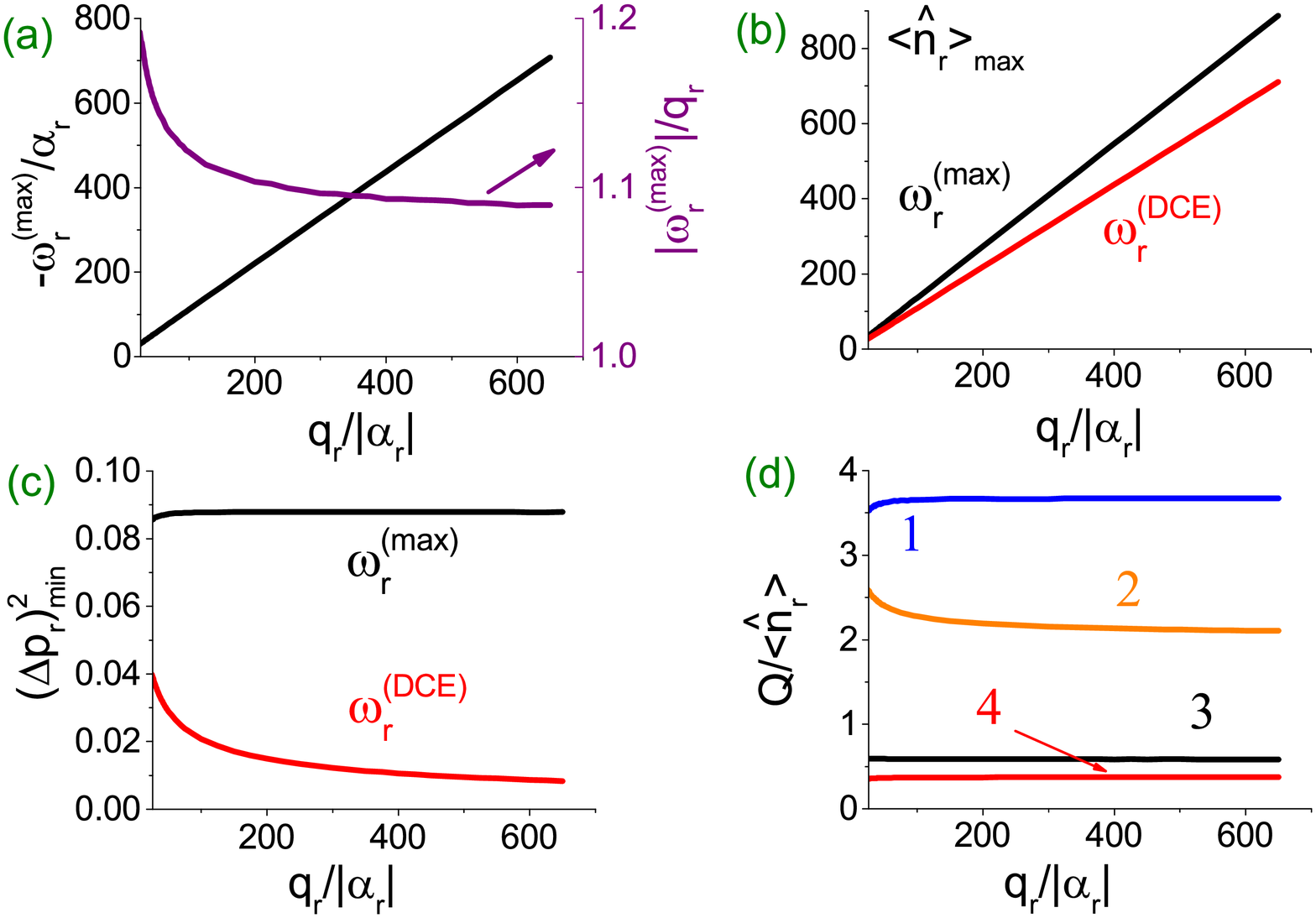} {}
\end{center}
\caption{\textbf{a)} Behavior of the effective frequency $%
\protect\omega _{r}^{(\max )}$ (that maximizes $\langle \hat{n}_{r}\rangle $%
) as function of $q_{r}/|\protect\alpha _{r}|$. \textbf{b)} Behavior of $%
\left\langle \hat{n}_{r}\right\rangle _{\max }$ as function of $q_{r}/|%
\protect\alpha _{r}|$ for the effective frequencies $\protect\omega %
_{r}^{(\max )}$ and $\protect\omega _{r}^{(\mathrm{DCE})}\equiv 0$. \textbf{%
c)} Behavior of $\left( \Delta p_{r}\right) _{\min }^{2}$ for these
effective frequencies. \textbf{d)} Behavior of $Q/\left\langle \hat{n}%
_{r}\right\rangle $ at different time instants. Curve 1 (2): $\protect\omega %
_{r}=\protect\omega _{r}^{(\max )}$ ($\protect\omega _{r}^{(\mathrm{DCE})}$)
and the time instant of the minimum value $\left( \Delta p_{r}\right) ^{2}$.
Curve 3 (4): $\protect\omega _{r}=\protect\omega _{r}^{(\max )}$ ($\protect%
\omega _{r}^{(\mathrm{DCE})}$) and the time instant of the maximum value of $%
\left\langle \hat{n}_{r}\right\rangle $.}
\label{fig4}
\end{figure}

For small ratio $q_{r}/|\alpha _{r}|\ll 1$ only two photons are generated
from vacuum for $\omega _{r}=-2\alpha _{r}$, as predicted by equation (\ref%
{wr}). For larger ratios $q_{r}/|\alpha _{r}|$ the behavior becomes much
more complicated and the dynamics is shown in figures \ref{fig2} and \ref%
{fig3} for different values of the effective detuning $\omega _{r}$ (which
can be adjusted experimentally by tuning the resonance shift $\zeta $). We
plot the time behavior of the average photon number $\langle \hat{n}%
_{r}\rangle $, the Mandel $Q$-factor and the variance of the squeezed field
quadrature $(\Delta p_{r})^{2}\equiv \langle \hat{p}_{r}^{2}\rangle -\langle
\hat{p}_{r}\rangle ^{2}$, where%
\begin{equation}
Q=\frac{\langle \hat{n}_{r}(\hat{n}_{r}-1)\rangle -\langle \hat{n}%
_{r}\rangle ^{2}}{\langle \hat{n}_{r}\rangle }~,~\hat{p}_{r}=\frac{\hat{a}%
_{r}-\hat{a}_{r}^{\dagger }}{2i}~.
\end{equation}%
In figure \ref{fig2} we set $q_{r}=3|\alpha _{r}|$ and in figure \ref{fig3} $%
q_{r}=650|\alpha _{r}|$. As expected, by increasing the ratio $q_{r}/|\alpha
_{r}|$ more photons are created from vacuum, and $\langle \hat{n}_{r}\rangle
$ can be optimized by choosing an appropriate value of $\omega _{r}$. The
average photon number is limited from above and exhibits a sort of irregular
collapse--revival behavior due to the Kerr nonlinearity, as opposed to the
exponential photon growth for the pure DCE case \cite%
{diss1,diss2,diss3,diss4,diss5,diss6}. The quantities $Q$ and $(\Delta
p_{r})^{2}$ also undergo oscillations, but they do not collapse to their
initial values, meaning that the field state never returns to the vacuum
state. The collapse--revival behavior of $\langle \hat{n}_{r}\rangle $ was
discovered more than two decades ago in a slightly different system -- the
pulsed parametric oscillator with a Kerr nonlinearity, where the classical
and quantum dynamics were compared \cite{milburn1,milburn2,milburn3}.

The field state becomes squeezed in the $\hat{p}_{r}$-quadrature for initial
times, but the squeezing disappears for larger times \cite{kerr1}, contrary
to the ideal DCE case when $(\Delta p_{r})^{2}$ decreases exponentially with
time \cite{pra}. In the presence of damping (shown by the line 5 in figure %
\ref{fig2}) the photon generation is still possible, but the oscillations of
$\langle \hat{n}_{r}\rangle $, $Q$ and $(\Delta p_{r})^{2}$, including the
collapse--revival behavior, disappear \cite{milburn2}. Moreover, the
asymptotic value of the $Q$-factor (shown by the dash-dotted line in figure %
\ref{fig2} and that can be calculated exactly \cite{kryuch1}) differs
substantially from its value during the transient, meaning that the field
state for initial times is quite different from the asymptotic state. We
also investigated how the dynamics is modified if the initial state is
slightly different from the vacuum state. This can occur in actual
experiments at finite temperature, so we considered the initial thermal
state with the average photon number $\bar{n}$, described by the density
operator $\hat{\rho}=\sum_{n=0}^{\infty }\rho _{n}|n\rangle \langle n|$, $%
\rho _{n}=\bar{n}^{n}/(\bar{n}+1)^{n+1}$. The dashed line in figure \ref%
{fig2} shows the dynamics for $\bar{n}=0.1$ in the absence of damping, which
should be compared with the line 3 calculated for the initial vacuum state.
We see that for initial times the differences are very small and the
oscillations of quantities $\langle \hat{n}_{r}\rangle $, $Q$ and $(\Delta
p_{r})^{2}$ persist. Therefore minor deviations of the initial state from
the vacuum do not pose a serious threat on the experimental verification of
the nonlinear DCE.

From figures \ref{fig2} and \ref{fig3} we see that certain values of $\omega
_{r}$ maximize $\langle \hat{n}_{r}\rangle $ for times $q_{r}t\leq 5$. We
denote this value by $\omega _{r}^{(\max )}$, noting that for another time
interval the value of $\omega _{r}$ that maximizes the average photon number
may be different. Since in actual implementations it might be difficult to
maintain external modulations for a long period of time, the choice $%
q_{r}t\leq 5$ seems appropriate to reflect the experimental reality. Figure %
\ref{fig4}a shows the behavior of $\omega _{r}^{(\max )}$ as function of $%
q_{r}/|\alpha _{r}|$: for large values of $q_{r}/|\alpha _{r}|$ it is
roughly given by $|\omega _{r}^{(\max )}|=1.09q_{r}$. The maximum number of
photons $\left\langle \hat{n}_{r}\right\rangle _{\max }$ when the effective
detuning is adjusted to $\omega _{r}^{(\max )}$ is shown in figure \ref{fig4}%
b: for $q_{r}/|\alpha _{r}|\gg 1$ it grows as $\left\langle \hat{n}%
_{r}\right\rangle _{\max }(\omega _{r}^{(\max )})=1.36q_{r}/|\alpha _{r}|$.
On the other hand, if we ignore this optimization and set the value of $%
\omega _{r}$ to the standard DCE resonance (without the Kerr nonlinearity), $%
\omega _{r}^{(\mathrm{DCE})}\equiv 0$, then $\left\langle \hat{n}%
_{r}\right\rangle _{\max }$ still grows linearly but with a smaller slope: $%
\left\langle \hat{n}_{r}\right\rangle _{\max }(\omega _{r}^{(\mathrm{DCE}%
)})=1.09q_{r}/|\alpha _{r}|$. Hence for large values of $q_{r}/|\alpha _{r}|$
the optimization can significantly enhance the photon generation,
facilitating the experimental verification. The downside of optimizing the
effective detuning to increase $\left\langle \hat{n}_{r}\right\rangle _{\max
}$ is that the squeezing is decreased. Figure \ref{fig4}c shows the smallest
value of $(\Delta p_{r})^{2}$ achieved for $q_{r}t\leq 5$ when $\omega _{r}$
is set to $\omega _{r}^{(\max )}$ or $\omega _{r}^{(\mathrm{DCE})}$.
Although in both cases the $\hat{p}_{r}$-quadrature becomes squeezed, for $%
\omega _{r}^{(\mathrm{DCE})}$ the squeezing is significantly stronger than
for $\omega _{r}^{(\max )}$.

Finally, in figure \ref{fig4}d we study the cavity field statistics at
different time instants by plotting the quantity $Q/\langle \hat{n}%
_{r}\rangle $ that quantifies the spread of the photon number distribution.
We recall that $Q/\langle \hat{n}_{r}\rangle =0$ for the coherent state, $%
Q/\langle \hat{n}_{r}\rangle =1$ for the thermal state and $Q/\langle \hat{n}%
_{r}\rangle =2+1/\langle \hat{n}_{r}\rangle $ for the squeezed vacuum state.
The states for which $Q/\langle \hat{n}_{r}\rangle >2+1/\langle \hat{n}%
_{r}\rangle $, called \textquotedblleft hyper-Poissonian\textquotedblright\
in \cite{hyper}, have photon number distributions distinguished by very long
tales with low probabilities that nonetheless cannot be neglected, so the
mean photon number does not characterizes well the total distribution \cite%
{2}. The curves 1 and 2 denote the value of $Q/\langle \hat{n}_{r}\rangle $
at the time instant of minimum $(\Delta p_{r})^{2}$ (shown in figure \ref%
{fig4}c) for $\omega _{r}^{(\max )}$ and $\omega _{r}^{(\mathrm{DCE})}$,
respectively. We see that for $\omega _{r}^{(\mathrm{DCE})}$ we have
approximately the squeezed vacuum state with $Q/\langle \hat{n}_{r}\rangle
\approx 2+1/\langle \hat{n}_{r}\rangle $, while for $\omega _{r}^{(\max )}$
we obtain a hyper-Poissonian state with a rather broad photon number
distribution. Lines 3 and 4 denote the value of $Q/\langle \hat{n}%
_{r}\rangle $ at the time instant of maximum $\langle \hat{n}_{r}\rangle $
(shown in figure \ref{fig4}b) for $\omega _{r}^{(\max )}$ and $\omega _{r}^{(%
\mathrm{DCE})}$, respectively. In this case the field states are not very
different one from another and have a super-Poissonian photon number
distribution with $0<Q/\langle \hat{n}_{r}\rangle <1$.

Summarizing, in the presence of the Kerr nonlinearity one can optimize the
photon generation from vacuum by adjusting the resonance shift $\zeta $
(directly related to the effective detuning $\omega _{r}$) as function of $%
\alpha _{r}$, and this is the second reason for the introduction of $\zeta $
in equation (\ref{eta}). On one hand, this optimization decreases the amount
of squeezing \cite{kerr1}, but on the other hand it can be used to produce
novel field states \cite{2} very different from the squeezed vacuum state
produced in standard DCE.

\subsection{External classical pumping}

If the cavity undergoes a classical pumping one must add the term $id\left(
\hat{a}^{\dagger }-\hat{a}\right) $ to the Hamiltonian (\ref{NHN}). In terms
of the auxiliary operators $\hat{A}$ and $\hat{B}$ we should add the term%
\begin{eqnarray}
\hat{H}_{d} &=&-\frac{1}{2}\sum\nolimits_{j}^{\prime }\varepsilon
_{d}^{(j)}e^{-it\left( \Delta _{+}/2-\tilde{\delta}_{+}-\delta _{\chi }-\eta
^{(j)}\right) }\left[ e^{-it\left( \Delta _{-}+2\tilde{\delta}_{-}\right)
/2}+\frac{\tilde{g}_{0}^{2}}{\Delta _{-}^{2}}e^{it\left( \Delta _{-}+2\tilde{%
\delta}_{-}\right) /2}\right] \hat{A}  \nonumber \\
&& \quad-\frac{\tilde{g}_{0}}{2\Delta _{-}}\sum\nolimits_{j}^{\prime }\varepsilon
_{d}^{(j)}e^{-it\left( \Delta _{+}/2-\tilde{\delta}_{+}-\eta ^{(j)}\right) }%
\left[ e^{-it\left( \Delta _{-}+2\tilde{\delta}_{-}\right) /2}-e^{it\left(
\Delta _{-}+2\tilde{\delta}_{-}\right) /2}\right] \hat{B}+h.c.\quad \quad
\end{eqnarray}%
to the RHS of Hamiltonian (\ref{vista}). To the lowest order in $\tilde{g}%
_{0}/\Delta _{-}$, for the pump frequency $\eta ^{(p)}=(\omega _{0}+\tilde{%
\delta}_{-}-\tilde{\delta}_{+}-\delta _{\chi }-\zeta )$ one can simply add
the effective pump Hamiltonian
\begin{equation}
\hat{H}_{p}=i\frac{1}{2}\left( \varepsilon _{d}^{(p)\ast }e^{i\phi _{q}/2}%
\hat{a}_{r}^{\dagger }-\varepsilon _{d}^{(p)}e^{-i\phi _{q}/2}\hat{a}%
_{r}\right)
\end{equation}%
to the RHS of equation (\ref{seth0}), where the operator $\hat{a}_{r}$ was
defined in equation (\ref{entre}).

Neglecting the Kerr nonlinearity, the optimum resonance shift for DCE is $%
\zeta =0$. For the simultaneous modulation of the system parameters and the
external pumping we obtain the general Hamiltonian of the form%
\begin{equation}
\hat{H}_{pump}\equiv \varrho \hat{a}_{r}+\frac{1}{2}\xi \hat{a}%
_{r}^{2}+h.c.~,  \label{v}
\end{equation}%
where we introduced arbitrary time-independent complex coefficients $\varrho $
and $\xi $. In the Heisenberg picture the solution for the Hamiltonian (%
\ref{v}) is straightforward:%
\begin{equation}
\hat{a}_{r}=\hat{a}_{r}(0)\cosh (\left\vert \xi \right\vert t)-i\frac{%
\xi ^{\ast }}{\left\vert \xi \right\vert }\hat{a}_{r}^{\dagger }(0)\sinh
(\left\vert \xi \right\vert t)+\frac{\varrho }{\xi }[\cosh (\left\vert
\xi \right\vert t)-1]-i\frac{\varrho ^{\ast }}{\left\vert \xi \right\vert
}\sinh (\left\vert \xi \right\vert t)~.
\end{equation}%
For the initial vacuum state, $\hat{a}_{r}|0\rangle =0$, we obtain for the
average photon number%
\begin{equation}
\left\langle \hat{n}_{r}\right\rangle =\sinh ^{2}(\left\vert \xi
\right\vert t)+2\frac{\left\vert \varrho \right\vert ^{2}}{\left\vert \xi
\right\vert ^{2}}\left[ \cosh (\left\vert \xi \right\vert t)-\sinh
(\left\vert \xi \right\vert t)\sin \left( 2\phi _{\varrho}-\phi _{\xi }\right) %
\right] \left[ \cosh (\left\vert \xi \right\vert t)-1\right] ~,
\end{equation}%
where we defined the phases as $\varrho =|\varrho |\exp (i\phi _{\varrho})$ and $%
\xi =|\xi |\exp (i\phi _{\xi })~.$

For initial times, $|\xi |t\ll 1$, we obtain%
\begin{equation}
\left\langle \hat{n}_{r}\right\rangle \approx ( \left\vert \xi \right\vert
^{2}+\left\vert \varrho \right\vert ^{2}) t^{2}~,
\end{equation}%
so the average photon number does not depend on the phases $\phi _{\varrho }$
and $\phi _{\xi }$. However, for larger times the phases become very
important as one gets%
\begin{equation}
\left\langle n\right\rangle =\sinh ^{2}(\left\vert \xi \right\vert t)+2%
\frac{\left\vert \varrho \right\vert ^{2}}{\left\vert \xi \right\vert ^{2}}%
e^{\pm \left\vert \xi \right\vert t}\left[ \cosh (\left\vert \xi
\right\vert t)-1\right] ,~~\mbox{for }2\phi _{\varrho}-\phi _{\xi }=\mp \frac{%
\pi }{2}+2\pi k~,
\end{equation}%
where $k$ is any integer number. In particular, for large times $|\xi
|t\gg 1$ we get%
\begin{equation}
\left\langle n\right\rangle \approx \left\{
\begin{array}{c}
\frac{1}{4}e^{2\left\vert \xi \right\vert t}+\frac{\left\vert \varrho
\right\vert ^{2}}{\left\vert \xi \right\vert ^{2}}e^{2\left\vert \xi
\right\vert t}~,~~\quad \mbox{for }2\phi _{\varrho}-\phi _{\xi }=-\frac{\pi }{2}%
+2\pi k \\
\frac{1}{4}e^{2\left\vert \xi \right\vert t}+\frac{\left\vert \varrho
\right\vert ^{2}}{\left\vert \xi \right\vert ^{2}}~, \quad\mbox{for }2\phi
_{\varrho}-\phi _{\xi }=\frac{\pi }{2}+2\pi k%
\end{array}%
\right. ~.
\end{equation}%
Therefore, by adjusting carefully the phase difference $(2\phi _{\varrho}-\phi
_{\xi })$ one can significantly amplify the photon generation with the
help of one-photon pumping. Moreover, one could verify our model
experimentally by measuring the dependence of $\left\langle \hat{n}%
_{r}\right\rangle $ on the phase either of the pump or the modulation
parameters $\varepsilon _{X}^{(j)}$ defined in equation (\ref{jones}).

\section{DCE-like behaviors with atomic clouds}

\label{clouds}

Besides forming the base of the toy model for Cavity DCE, for $N\gg 1$ the
Hamiltonian (\ref{H1}) also describes accurately the interaction between a
cold \textquotedblleft atomic cloud\textquotedblright\ (e.g., polar
molecules \cite{polar1,polar2}) or ensemble of superconducting qubits and a
single cavity mode. In this case all the parameters of Hamiltonian (\ref{NHN}%
) are controllable, and novel regimes of light--matter interaction can be
implemented by modulating the system externally according to the law of
motion (\ref{jones}). The full solution is given in \ref{an}, and in this
section we highlight the regimes in which excitations can be generated from
vacuum. In section \ref{secADCE} we shall describe another regime when pair
of excitations can be coherently annihilated due to external modulation, in
what we call \textquotedblleft Anti-DCE\textquotedblright .

In the dispersive regime, for the modulation frequency%
\begin{equation}
\eta ^{(I)}=2(\Omega _{0}-\tilde{\delta}_{-}-\tilde{\delta}_{+}-\zeta )~,
\end{equation}%
we obtain the effective Hamiltonian%
\begin{equation}
\hat{H}_{eff}=\left[ e^{-2it\zeta }q_{I}\left( \hat{B}^{2}-2e^{it\delta
_{\chi }}\frac{\tilde{g}_{0}}{\Delta _{-}}\hat{A}\hat{B}+e^{2it\delta _{\chi
}}\frac{\tilde{g}_{0}^{2}}{\Delta _{-}^{2}}\hat{A}^{2}\right) +h.c.\right] +%
\hat{H}_{NG}~,  \label{jason}
\end{equation}%
where the non-Gaussian part $\hat{H}_{NG}$ is given by equations (\ref{IDCE1}%
) and (\ref{IDCE2}) and we defined the time-independent parameter%
\begin{equation}
q_{I}=-\frac{\tilde{\delta}_{-}}{2}\left[ \left( i-\frac{2\chi _{0}}{\Delta
_{-}}\right) \frac{\varepsilon _{\omega }^{(I)}}{\Delta _{+}}+i\frac{2\omega
_{0}}{\Delta _{+}}\left( \frac{\varepsilon _{\Omega }^{(I)}}{2\Omega _{0}}-%
\frac{\tilde{\varepsilon}_{g}^{(I)}}{\tilde{g}_{0}}\right) +\frac{%
\varepsilon _{\chi }^{(I)}}{\Delta _{-}}\right] ~.
\end{equation}%
Neglecting the non-linear terms and considering $|\delta _{\chi }|\ll
|q_{I}| $, we can write $\hat{A}(t)\simeq \hat{A}(0)-e^{-it\delta _{\chi }}(%
\tilde{g}_{0}/\Delta _{-})[\hat{B}(t)-\hat{B}(0)]$, so to the lowest order
in $\tilde{g}_{0}/\Delta _{-}$ we obtain the total effective Hamiltonian
(for the initial zero-excitation state)%
\begin{equation}
\hat{H}_{eff}\simeq (q_{I}e^{-2it\zeta }\hat{B}^{2}+h.c.)+\delta _{-}[(\hat{B%
}^{\dagger }B)^{2}-\hat{B}^{\dagger }\hat{B}]~.  \label{ig}
\end{equation}%
Hamiltonian (\ref{ig}) is analogous to the DCE Hamiltonian (\ref{marty}) but
with the matter operator $\hat{B}$ instead of the cavity operator $\hat{A}$.
So this behavior corresponds to the DCE with matter, when pairs of atomic
internal excitations are created from vacuum instead of photons (recall that
the CM motion of atoms is prescribed externally). Notice that there is
analogous Kerr nonlinearity term $\delta _{-}(\hat{B}^{\dagger }B)^{2}$, yet
for $|q_{I}|\gg |\delta _{-}|$ many matter excitations can be created from
vacuum. We call this behavior \textquotedblleft \emph{Inverse dynamical
Casimir effect}\textquotedblright\ (IDCE), since figuratively this
phenomenon corresponds to exciting the internal degrees of freedom of the
moving dielectric slab instead of creating photons. In section \ref{secADCE}
we shall also describe the \textquotedblleft Anti-IDCE\textquotedblright\
phenomenon -- an analog of Anti-DCE for the atomic degrees of freedom.

For some modulation frequencies one can achieve simultaneous excitation of
the cavity and the atoms. In the dispersive regime this occurs for the
modulation frequency $\eta ^{(M)}=\Delta _{+}-2\tilde{\delta}_{+}-\delta
_{\chi }-\zeta $, when the effective Hamiltonian reads%
\begin{equation}
\hat{H}_{eff}=\left[ q_{M}e^{-it\zeta }\left( -\hat{A}\hat{B}+e^{it\delta
_{\chi }}\frac{\tilde{g}_{0}}{\Delta _{-}}\hat{A}^{2}-e^{-it\delta _{\chi }}%
\frac{\tilde{g}_{0}}{\Delta _{-}}\hat{B}^{2}\right) +h.c.\right] +\hat{H}%
_{NG}(\tilde{g}_{0})+\hat{H}_{NG}(\tilde{\varepsilon}_{g})~.  \label{pari}
\end{equation}%
The non-Gaussian parts are given by equations (\ref{IDCE1}) -- (\ref{MDCE})
and%
\begin{equation}
q_{M}=\mathcal{-}\frac{\tilde{g}_{0}}{2}\left[ \left( i-\frac{4\chi _{0}}{%
\Delta _{-}}\right) \frac{\varepsilon _{\omega }^{(M)}}{\Delta _{+}}+i\frac{%
\varepsilon _{\Omega }^{(M)}}{\Delta _{+}}-i\frac{\tilde{\varepsilon}%
_{g}^{(M)}}{\tilde{g}_{0}}+2\frac{\varepsilon _{\chi }^{(M)}}{\Delta _{-}}%
\right] ~.
\end{equation}%
We call this behavior \textquotedblleft Mixed behavior\textquotedblright ,
since the photons and atomic excitations are created at the same rate $%
|q_{M}|\gg |q_{I}|$.

In the resonant regime, $\Delta _{-}=0$, excitations are generated from
vacuum for the modulation frequency $\eta ^{(R)}=2\omega _{0}-2\tilde{\delta}%
_{+}-\delta _{\chi }-\zeta $. The total effective Hamiltonian is%
\begin{eqnarray}
\hat{H}_{eff} &=&\tilde{g}_{0}e^{-it\zeta }\left[ \left( \Theta
_{+}^{(R)}e^{-2it\tilde{g}_{0}}-\Theta _{-}^{(R)}e^{2it\tilde{g}_{0}}\right)
\hat{A}\hat{B}\right.  \nonumber \\
&&+\frac{1}{2}e^{it\delta _{\chi }}\left( \Theta _{0}^{(R)}+\Theta
_{+}^{(R)}e^{-2it\tilde{g}_{0}}+\Theta _{-}^{(R)}e^{2it\tilde{g}_{0}}\right)
\hat{A}^{2}  \nonumber \\
&&\left. +\frac{1}{2}e^{-it\delta _{\chi }}\left( -\Theta _{0}^{(R)}+\Theta
_{+}^{(R)}e^{-2it\tilde{g}_{0}}+\Theta _{-}^{(R)}e^{2it\tilde{g}_{0}}\right)
\hat{B}^{2}+h.c.\right]  \nonumber \\
&&-\frac{\tilde{g}_{0}}{8N}\left( \hat{B}^{\dagger }\hat{B}+\hat{A}^{\dagger
}\hat{A}-1\right) \left( \hat{A}\hat{B}^{\dagger }e^{it\delta _{\chi }}+\hat{%
A}^{\dagger }\hat{B}e^{-it\delta _{\chi }}\right) ~.  \label{po}
\end{eqnarray}%
The time-independent coefficients $\Theta _{i}^{(R)}$ are given by equation (%
\ref{guest}) and we neglected the non-Gaussian terms proportional to $\tilde{%
\varepsilon}_{g}$. We see that for $\zeta =0,\pm 2\tilde{g}_{0}$ one can
create equal amounts of cavity and matter excitations. The photon generation
for $\zeta =0$ was unknown until a few years ago \cite{sinaia1,sinaia2}, yet
it appears naturally in our formalism, as well as the non-Gaussian terms on
the last line of equation (\ref{po}). The detailed analysis of Hamiltonian (%
\ref{NHN}) in the resonant regime (without the non-linear terms) was studied
in \cite{pla,pra,pla1,pla2,pla3} in an attempt to describe the detection of
DCE using small induction loops modeled as LC contours.

\section{Nonstationary circuit QED with a single qubit}

\label{qubit}

Now we consider the limiting case $N=1$ to study which phenomena exist for
the most basic type of light--matter interaction under nonstationary
conditions. From the practical point of view this analysis is relevant
because it describes actual implementations in the circuit QED architecture,
where the parameters of the cavity and the qubit can be modulated \emph{in
situ} by external biases and the one-photon classical pump is implemented in
a straightforward manner \cite{cir1,cir2,cir3,meta,nori-n}. Nonstationary
circuit QED has been studied in numerous papers during the last decade \cite%
{me-arx,jpcs,1,liberato9,2level,2,2atom,3level,sinaia1,CAMOP-me}, but here
we generalize the previous results by working in the dressed-states basis
\cite{JPA} and considering the simultaneous multi-tone modulation of all the
system parameters. Moreover, we predict the new effect in which pair of
excitations can be coherently annihilated due to external modulation, in
what we call \textquotedblleft Anti-DCE\textquotedblright\ behavior.

As shown in \ref{circuitqed} the wavefunction corresponding to the
Hamiltonian (\ref{H1}) can be written approximately as%
\begin{equation}
|\psi (t)\rangle \simeq e^{-it\bar{\lambda}_{0}}b_{0}(t)|\varphi _{0}\rangle
+\sum_{n=1}^{\infty }\sum_{\mathcal{S}=\pm }e^{-it\bar{\lambda}_{n,\mathcal{S%
}}}b_{n,\mathcal{S}}(t)|\varphi _{n,\mathcal{S}}\rangle ~.
\end{equation}%
Here $|\varphi _{n,\mathcal{S}}\rangle $ and $\bar{\lambda}_{n,\mathcal{S}}$
are the $n$-excitations eigenstates (also known as \emph{dressed states})
and the \textquotedblleft corrected\textquotedblright\ eigenvalues of the
bare Jaynes-Cummings Hamiltonian
\begin{equation}
\hat{H}_{JC}=\omega _{0}\hat{n}+\Omega _{0}|e\rangle \langle e|+g_{0}(\hat{a}%
\hat{\sigma}_{+}+\hat{a}^{\dagger }\hat{\sigma}_{-})~.
\end{equation}%
Coefficients $b$ represent approximately the probability amplitudes of the
dressed states and the index $\mathcal{S}$ labels the different eigenstates
with the same number of excitations. The corrected eigenfrequencies and the
eigenstates read approximately%
\begin{equation}
\bar{\lambda}_{0}\simeq 0~,~\bar{\lambda}_{n>0,\mathcal{S}}\simeq \omega
_{0}n-\frac{\Delta _{-}}{2}+\mathcal{S}\frac{1}{2}\beta _{n}~,~\beta _{n}=%
\sqrt{\Delta _{-}^{2}+4g_{0}^{2}n}~,~\mathcal{S}=\pm ~
\end{equation}%
\begin{equation}
|\varphi _{0}\rangle =|g,0\rangle ~,~|\varphi _{n>0,\mathcal{S}}\rangle ={\rm s}%
_{n,\mathcal{S}}|g,n\rangle +{\rm c}_{n,\mathcal{S}}|e,n-1\rangle ~,
\end{equation}%
where $\Delta _{\pm }=\omega _{0}\pm \Omega _{0}$ and we introduced the
notation%
\begin{equation}
{\rm s}_{m,+}=\sin \theta _{m},~{\rm s}_{m,-}=\cos \theta _{m},~{\rm c}_{m,+}=\cos
\theta _{m},~{\rm c}_{m,-}=-\sin \theta _{m}~
\end{equation}%
\begin{equation}
\theta _{m>0}=\arctan \frac{\Delta _{-}+\beta _{m}}{2g_{0}\sqrt{m}}~.
\end{equation}

\subsection{DCE behavior}

For a single modulation frequency matching the DCE resonance, $\eta
^{(D)}\approx 2\omega _{0}$, the probability amplitudes obey the
differential equation [see equations (\ref{b}) -- (\ref{ap3})]%
\begin{equation}
\dot{b}_{m,\mathcal{T}}=\sum_{\mathcal{S}}[\Theta _{m+2,\mathcal{T},\mathcal{%
S}}^{(D)}e^{-it[\bar{\lambda}_{m+2,\mathcal{S}}-\bar{\lambda}_{m,\mathcal{T}%
}-\eta ^{(D)}]}b_{m+2,\mathcal{S}}-\Theta _{m,\mathcal{S},\mathcal{T}%
}^{(D)\ast }e^{it[\bar{\lambda}_{m,\mathcal{T}}-\bar{\lambda}_{m-2,\mathcal{S%
}}-\eta ^{(D)}]}b_{m-2,\mathcal{S}}]~,
\end{equation}%
where we use the shorthand notation $b_{0,\mathcal{T}}\equiv b_{0}$, $\bar{%
\lambda}_{0,\mathcal{T}}\equiv \bar{\lambda}_{0}$; $\Theta _{m+2,\mathcal{T},%
\mathcal{S}}^{(D)}$ is a time-independent coefficient given by equations (%
\ref{teta1}) and (\ref{teta2}). In the argument of the exponential functions
there is an intrinsic uncertainty we call \textquotedblleft Systematic-error
frequency shift\textquotedblright\ (SEFS) $\nu _{m,\mathcal{T}}^{(3)}$ due
to the involved approximations. The estimative of $\nu _{m,\mathcal{T}%
}^{(3)} $ is given in \ref{circuitqed}, since its order of magnitude is
important to tune precisely the resonant modulation frequency.

The frequency $\eta ^{(D)}$ matches the difference $\bar{\lambda}_{m+2,%
\mathcal{S}}-\bar{\lambda}_{m,\mathcal{T}}$ only in the \emph{dispersive
regime}, $|\Delta _{-}|/2\gg g_{0}\sqrt{n}$, where $n$ is an integer
representing the number of excitations. Introducing the \textquotedblleft
detuning symbol\textquotedblright\ $\mathcal{D}\equiv \Delta _{-}/|\Delta
_{-}|=\pm $ we can write the corrected eigenvalues and eigenstates as (we
denote $\bar{\lambda}_{0}\equiv \bar{\lambda}_{0,\mathcal{D}}$)%
\begin{equation}
\bar{\lambda}_{m,\mathcal{D}}=\omega _{g}m-\alpha m^{2}-\delta _{+}-\frac{%
\delta _{\chi }}{2}~,~\bar{\lambda}_{m>0,\mathcal{-D}}=~\omega
_{e}m-\allowbreak \Delta _{-}+\alpha m^{2}-\delta _{+}+\frac{\delta _{\chi }%
}{2}
\end{equation}%
\begin{equation}
|\varphi _{m,\mathcal{D}}\rangle \simeq |\mathbf{g},m\rangle +\frac{g_{0}%
\sqrt{m}}{\Delta _{-}}|\mathbf{e},m-1\rangle ~,~|\varphi _{m>0,\mathcal{-D}%
}\rangle \simeq -\mathcal{D}(|\mathbf{e},m-1\rangle -\frac{g_{0}\sqrt{m}}{%
\Delta _{-}}|\mathbf{g},m\rangle )~,
\end{equation}%
where the effective cavity frequency $\omega _{g}$ or $\omega _{e}$ and the
intrinsic \textquotedblleft frequency shifts\textquotedblright\ are%
\begin{equation}
\omega _{g}\equiv \omega _{0}+\allowbreak \delta _{-}-\delta _{+}-\delta
_{\chi }~,~\omega _{e}\equiv \omega _{0}-\delta _{-}+\delta _{+}-\delta
_{\chi }~,~\delta _{\pm }\equiv \frac{g_{0}^{2}}{\Delta _{\pm }}~,~\delta
_{\chi }\equiv \frac{4\chi _{0}^{2}}{\Delta _{+}}.
\end{equation}%
Hence the modulation frequency $\eta ^{(D)}$ can couple either the dressed
states $|\varphi _{m,\mathcal{D}}\rangle \leftrightarrow |\varphi _{m\pm 2,%
\mathcal{D}}\rangle $ (where $|\varphi _{0,\mathcal{D}}\rangle \equiv
|\varphi _{0}\rangle $) or $|\varphi _{m>0,-\mathcal{D}}\rangle
\leftrightarrow |\varphi _{m\pm 2,-\mathcal{D}}\rangle $. The former case
occurs when the atom is predominantly in the ground state, so we call it
\emph{g-DCE behavior}. The second case corresponds to the atom predominantly
in the excited state, so we call it \emph{e-DCE behavior}.

\emph{1) g-DCE behavior.} Under the approximations and SEFS%
\begin{equation}
|\Theta _{m+2,-\mathcal{D},\mathcal{D}}^{(D)}|,|\Theta _{m+2,\mathcal{D},%
\mathcal{-D}}^{(D)}|\ll |\Delta _{-}|,~|\Theta _{m+2,-\mathcal{D},\mathcal{-D%
}}^{(D)}|\ll |\delta _{-}|  \label{paso1}
\end{equation}%
\begin{equation}
O(\nu _{m,\mathcal{T}}^{(3)})\sim |\Theta _{m+2,-\mathcal{D},\mathcal{D}%
}^{(D)}|^{2}/|\Delta _{-}|,~|\Theta _{m+2,\mathcal{D},\mathcal{-D}%
}^{(D)}|^{2}/|\Delta _{-}|,~|\Theta _{m+2,-\mathcal{D},\mathcal{-D}%
}^{(D)}|^{2}/|\delta _{-}|  \label{paso2}
\end{equation}%
we define the effective probability amplitudes $c_{m}$ as%
\begin{equation}
c_{m}=\left\{
\begin{array}{c}
b_{0}~,~m=0 \\
b_{m,\mathcal{D}}~,~m>0%
\end{array}%
\right. ~.
\end{equation}%
Adjusting the modulation frequency to%
\begin{equation}
\eta ^{(D)}=2(\omega _{g}-\zeta )~,  \label{fre1}
\end{equation}%
where $\zeta $ is an adjustable resonance shift, we obtain the differential
equations%
\begin{eqnarray}
\dot{c}_{m} &=&\left\vert \vartheta _{+}\right\vert \left[ e^{i\phi _{+}}%
\sqrt{(m+1)(m+2)}e^{-2it\left[ \zeta -2\alpha (m+1)\right] }c_{m+2}\right.
\nonumber \\
&&\left. -e^{-i\phi _{+}}\sqrt{m(m-1)}e^{2it\left[ \zeta -2\alpha (m-1)%
\right] }c_{m-2}~\right]
\end{eqnarray}%
\begin{equation}
\vartheta _{+}\equiv \frac{1}{2}\left[ \left( \frac{\Omega _{0}}{\Delta _{+}}%
-i\frac{\chi _{0}}{\delta _{-}}\right) \delta _{-}\frac{\varepsilon _{\omega
}^{(D)}}{\omega _{0}}+\delta _{-}\frac{\varepsilon _{\Omega }^{(D)}}{\Delta
_{+}}-\frac{2\delta _{-}\Omega _{0}}{\Delta _{+}}\frac{\varepsilon _{g}^{(D)}%
}{g_{0}}+i\varepsilon _{\chi }^{(D)}\right] =\left\vert \vartheta
_{+}\right\vert e^{i\phi _{+}}~.  \label{vi1}
\end{equation}%
Comparing with equations (\ref{seth0}) and (\ref{seth}) we see that under a
trivial phase rotation $\hat{a}_{r}\rightarrow \hat{a}_{r}e^{i(\phi _{+}+\pi
)/2}$ the dynamics of $c_{m}$ is described by the nonlinear DCE Hamiltonian (%
\ref{seth0}) with $\omega _{r}=\zeta $, $\alpha _{r}=-\alpha $ and $%
q_{r}=\left\vert \vartheta _{+}\right\vert $. If we also apply a one-photon
pump with the frequency $\eta ^{\left( p\right) }=\omega _{g}-\zeta $, then
one can simply add the term $-[\varepsilon _{d}^{(p)}e^{-i(\phi _{+}+\pi )/2}%
\hat{a}_{r}/2+h.c.]$ to the RHS of equation (\ref{seth0}) under the
additional approximations%
\begin{equation}
\frac{\varepsilon _{d}\sqrt{m}}{|\delta _{-}|},\frac{\varepsilon _{d}g_{0}}{%
\Delta _{-}^{2}}\ll 1~,\quad ~O(\nu _{m,\mathcal{T}}^{(3)})\sim \frac{%
\varepsilon _{d}^{2}}{|\delta _{-}|}~.  \label{hell}
\end{equation}

\emph{2) e-DCE behavior.} On the other hand, for the modulation frequency
\begin{equation}
\eta ^{(D)}=2(\omega _{e}-\zeta )  \label{fre2}
\end{equation}%
we define $c_{m}=b_{m+1,\mathcal{-D}}$ and obtain the differential equations%
\begin{eqnarray}
\dot{c}_{m} &=&\left\vert \theta _{-}\right\vert \left[ e^{i\phi _{-}}\sqrt{%
(m+1)(m+2)}e^{-2it[\zeta +2\alpha (m+2)]}c_{m+2}\right.  \nonumber \\
&&\left. -e^{-i\phi _{-}}\sqrt{m(m-1)}e^{2it[\zeta +2\alpha m]}c_{m-2}\right]
~,
\end{eqnarray}%
where%
\begin{equation}
\vartheta _{-}\equiv \frac{1}{2}\left[ -\left( \frac{\Omega _{0}}{\Delta _{+}%
}+i\frac{\chi _{0}}{\delta _{-}}\right) \delta _{-}\frac{\varepsilon
_{\omega }^{(D)}}{\omega _{0}}-\delta _{-}\frac{\varepsilon _{\Omega }^{(D)}%
}{\Delta _{+}}+\frac{2\delta _{-}\Omega _{0}}{\Delta _{+}}\frac{\varepsilon
_{g}^{(D)}}{g_{0}}+i\varepsilon _{\chi }^{(D)}\right] =\left\vert \vartheta
_{-}\right\vert e^{i\phi _{-}}~.  \label{vi2}
\end{equation}%
These results are valid under the approximations (\ref{paso1}) -- (\ref%
{paso2}) with replacement $\mathcal{D}\rightarrow -\mathcal{D}$. So the
dynamics of $c_{m}$ is again described by the nonlinear DCE Hamiltonian with
$\omega _{r}=\zeta +2\alpha $, $\alpha _{r}=\alpha $ and $q_{r}=|\vartheta
_{-}|$, and for the external pump with frequency $\eta ^{(p)}=\omega
_{e}-\zeta $ one can simply add the term $-[\varepsilon _{d}^{(p)}e^{-i(\phi
_{-}+\pi )/2}\hat{a}_{r}/2+h.c.]$.

Thus the nonlinear dynamical Casimir effect exists even for a singe qubit,
so it is an intrinsic phenomenon of the light--matter interaction in
nonstationary systems and can be observed in the circuit QED architecture.
For a single qubit there are two possible modulation frequencies, equations (%
\ref{fre1}) and (\ref{fre2}), whereas for $N\gg 1$ we found only one
resonant modulation frequency, equation (\ref{eta}). The origin of this
apparent discrepancy is trivial: in section \ref{toy} we assumed that $%
\langle \hat{b}^{\dagger }\hat{b}\rangle \ll N$, so the case when the atoms
were initially in the excited states was automatically excluded from the
treatment. The photon generation rates $\left\vert \vartheta _{+}\right\vert
$ and $\left\vert \vartheta _{-}\right\vert $ are of the same order of
magnitude, but may differ due to the phases $\phi _{X}^{(D)}$ for
simultaneous modulation of several parameter. The effective detuning $\omega
_{r}$ and the Kerr coefficient $\alpha _{r}$ depend on the initial atomic
states, so the resonance shift $\zeta $ must be adjusted accordingly to
optimize the photon generation. By increasing the number of qubits we simply
make the replacements $|\alpha |\rightarrow N|\alpha |$ and $\delta
_{-}\rightarrow N\delta _{-}$, as can be seen from equations (\ref{vi3}), (%
\ref{vi1}) and (\ref{vi2}). So for $\varepsilon _{\chi }=0$ the maximum
number of photons created from the initial zero-excitation state (shown in
figure \ref{fig4}b) is not altered by increasing the number of atoms,
although the photon generation rate undergoes a $N$-fold increase.

\subsection{Anti-DCE behavior}

\label{secADCE}

In the dispersive regime, for the modulation frequency (we neglect the Kerr
nonlinearity $\alpha $ to simplify the expressions)%
\begin{equation}
\eta _{M}^{(A)}=2\omega _{0}+\Delta _{-}-3\delta _{\chi }+2\allowbreak
\left( \delta _{-}-\delta _{+}\right) \left( M+1\right) ~,  \label{fr}
\end{equation}%
where $M$ is a positive integer, we obtain the differential equations for $%
m>0$%
\begin{eqnarray}
\dot{b}_{m,\mathcal{-D}} &=&\Theta _{m+2,\mathcal{-D},\mathcal{D}%
}^{(A)}e^{-2it\allowbreak \left( \delta _{-}-\delta _{+}\right) \left(
m-M\right) }b_{m+2,\mathcal{D}}  \nonumber \\
\dot{b}_{m+2,\mathcal{D}} &=&-\Theta _{m+2,\mathcal{-D},\mathcal{D}%
}^{(A)\ast }e^{2it\allowbreak \left( \delta _{-}-\delta _{+}\right) \left(
m-M\right) }b_{m,\mathcal{-D}}~  \label{texas}
\end{eqnarray}%
\begin{equation}
\Theta _{m+2,\mathcal{-D},\mathcal{D}}^{\left( A\right) }=\mathcal{D}\frac{%
\delta _{-}\Omega _{0}g_{0}}{2\omega _{0}\Delta _{-}}\sqrt{m(m+1)(m+2)}\left[
\frac{\varepsilon _{\omega }^{(A)}}{2\omega _{0}+\Delta _{-}}+\frac{\omega
_{0}+\Delta _{-}}{2\omega _{0}+\Delta _{-}}\frac{\varepsilon _{\Omega }^{(A)}%
}{\Omega _{0}}-\frac{\varepsilon _{g}^{(A)}}{g_{0}}\right] .
\end{equation}%
The involved approximations are%
\begin{equation}
|\Theta _{m+2,\mathcal{-D},\mathcal{-D}}^{\left( A\right) }|,|\Theta _{m+2,%
\mathcal{D},\mathcal{D}}^{\left( A\right) }|,|\Theta _{m+2,\mathcal{D},%
\mathcal{-D}}^{\left( A\right) }|\ll |\Delta _{-}|
\end{equation}%
\begin{equation}
O(\nu _{m,\mathcal{T}}^{(3)})\sim |\Theta _{m+2,\mathcal{-D},\mathcal{-D}%
}^{\left( A\right) }|^{2}/|\Delta _{-}|,~|\Theta _{m+2,\mathcal{D},\mathcal{D%
}}^{\left( A\right) }|^{2}/|\Delta _{-}|,~|\Theta _{m+2,\mathcal{D},\mathcal{%
-D}}^{\left( A\right) }|^{2}/|\Delta _{-}|~.
\end{equation}

Under realistic conditions we have $|\delta _{-}|\gg |\Theta _{M+2,-\mathcal{%
D},\mathcal{D}}^{(A)}|$, so only the amplitudes $b_{M,-\mathcal{D}}$ and $%
b_{M+2,\mathcal{D}}$ are effectively coupled. Therefore this modulation
roughly couples the states $|e,M-1\rangle \leftrightarrow |g,M+2\rangle $.
In other words, for the initial state $|g\rangle \otimes \sum_{m=0}^{\infty
}\rho _{m}|m\rangle $ one can couple the subsets $|g,M\rangle
\leftrightarrow |e,M-3\rangle $, thereby annihilating three photons (two
excitations in total) via external modulation. However the coupling rate $%
|\Theta _{m+2,\mathcal{-D},\mathcal{D}}^{(A)}|$ is very small, so the
frequency (\ref{fr}) must be fine tuned (taking into account the Kerr
nonlinearity and SEFS) and the transfer of populations between the states
takes a long time. If only $\varepsilon _{\chi }\neq 0$, as in parametric
down-conversion, then $\Theta _{m+2,\mathcal{-D},\mathcal{D}}^{\left(
A\right) }=0$ and this process does not occur at all. Noticing that in the
dispersive regime the coupling $|g,m\rangle \leftrightarrow |e,m-2\rangle $
via one-photon pumping is prohibited, we conclude that the subtraction of
system excitations via external modulation (when the atom starts in the
ground state) only occurs for the time-modulation of parameters $\omega $, $%
\Omega $ or $g$.

This phenomenon persists in the macroscopic case for a cold atomic cloud. As
shown in \ref{dispersive}, for the modulation frequency $\eta ^{(A)}\approx
2\omega _{0}+\Delta _{-}=3\omega _{0}-\Omega _{0}$ we obtain effective
Hamiltonian of the form $\hat{H}_{eff}\simeq \lbrack i\tilde{\delta}_{-}%
\tilde{\varepsilon}_{g}^{(A)}/(4N\Delta _{-})](\hat{A}^{3}\hat{B}^{\dagger }-%
\hat{A}^{\dagger 3}\hat{B})+\cdots $, where other nonlinear terms are given
by equations (\ref{IDCE1}) and (\ref{ADCE}) and we neglected the
contributions of $\varepsilon _{\omega }$ and $\varepsilon _{\Omega }$. So
when the atoms start in the ground states there is an annihilation of three
photons accompanied by generation of one collective atomic excitation. Since
in this case the photons are annihilated by virtue of external modulation of
the system parameters, including the prescribed motion of the atomic cloud,
we call this effect \textquotedblleft Anti-DCE\textquotedblright . This name
should not be taken too literally because such behavior cannot be
implemented with a dielectric slab for which the parameters $\Omega $ and $g$%
, and hence the resonant modulation frequency $\eta ^{(A)}$, are not known.

In section \ref{clouds} we described the IDCE behavior, when pairs of atomic
excitations are generated from vacuum for the modulation frequency $\eta
^{(I)}\approx 2\Omega _{0}$. By symmetry in Hamiltonian (\ref{NHN}), there
is also the \textquotedblleft Anti-IDCE\textquotedblright\ behavior, when
three atomic excitations are annihilated (for the cavity field in the ground
state) due to the modulation of system parameters with frequency $\eta
^{(AI)}\approx 2\Omega _{0}-\Delta _{-}=3\Omega _{0}-\omega _{0}$. This
effect is described by the effective Hamiltonian $\hat{H}_{eff}\simeq -[i%
\tilde{\delta}_{-}\tilde{\varepsilon}_{g}^{(AI)}/(4N\Delta _{-})](\hat{B}^{3}%
\hat{A}^{\dagger }-\hat{B}^{\dagger 3}\hat{A})+\cdots $, as given by
equations (\ref{IDCE1}) and (\ref{AIDCE}). In practice the Anti-DCE and
Anti-IDCE behaviors are very difficult to observe because the involved
coupling rates are quite small. However, they are interesting from the
purely theoretical point of view for constituting examples of motion-induced
coherent annihilation of excitations.

\subsection{Generation of entangled states}

Now we briefly review some practical schemes to generate entangled states in
circuit QED with time-modulated parameters, studied previously in \cite%
{me-arx,jpcs,1}. In the dispersive regime, for the modulation frequency
(neglecting the nonlinearity $\alpha $)
\begin{equation}
\eta _{M}^{(S)}=\Delta _{+}-2\left( \delta _{-}-\delta _{+}\right) \left(
M+1\right) -\delta _{\chi }
\end{equation}%
and approximations%
\begin{equation}
|\Theta _{m+2,\mathcal{D},\mathcal{D}}^{\left( S\right) }|,|\Theta _{m+2,-%
\mathcal{D},\mathcal{D}}^{\left( S\right) }|,|\Theta _{m+2,-\mathcal{D},-%
\mathcal{D}}^{\left( S\right) }|\ll |\Delta _{+}|~,~
\end{equation}%
\begin{equation}
O(\nu _{m,\mathcal{T}}^{(3)})\sim |\Theta _{m+2,\mathcal{D},\mathcal{D}%
}^{\left( S\right) }|^{2}/|\Delta _{+}|,~|\Theta _{m+2,-\mathcal{D},\mathcal{%
D}}^{\left( S\right) }|^{2}/|\Delta _{+}|,~|\Theta _{m+2,-\mathcal{D},-%
\mathcal{D}}^{\left( S\right) }|^{2}/|\Delta _{+}|
\end{equation}%
we obtain the equations (denoting $b_{0,\mathcal{D}}\equiv b_{0}$)%
\begin{eqnarray}
\dot{b}_{m,\mathcal{D}} &=&\Theta _{m+2,\mathcal{D},\mathcal{-D}%
}^{(S)}e^{-it2\left( \delta _{-}-\delta _{+}\right) \left( M-m\right)
}b_{m+2,\mathcal{-D}}  \nonumber \\
\dot{b}_{m+2,\mathcal{-D}} &=&-\Theta _{m+2,\mathcal{D},\mathcal{-D}%
}^{(S)\ast }e^{it2\left( \delta _{-}-\delta _{+}\right) \left( M-m\right)
}b_{m,\mathcal{D}}  \label{paris}
\end{eqnarray}%
\begin{equation}
\Theta _{m+2,\mathcal{D},\mathcal{-D}}^{\left( S\right) }=\frac{1}{2}g_{0}%
\mathcal{D}\sqrt{m+1}\left[ -\left( 1+\frac{4i\chi _{0}}{\Delta _{-}}\right)
\frac{\varepsilon _{\omega }^{(S)}}{\Delta _{+}}-\frac{\varepsilon _{\Omega
}^{(S)}}{\Delta _{+}}+\frac{\varepsilon _{g}^{(S)}}{g_{0}}+i\frac{%
2\varepsilon _{\chi }^{(S)}}{\Delta _{-}}\right] ~.
\end{equation}

For $|\delta _{-}|\gg |\Theta _{M+2,\mathcal{D},\mathcal{-D}}^{(S)}|$ only
the amplitudes $b_{M,\mathcal{D}}$ and $b_{M+2,\mathcal{-D}}$ are
effectively coupled, so this frequency roughly couples the states $%
|g,M\rangle \leftrightarrow |e,M+1\rangle $. For this reason such behavior
was called \textquotedblleft AJC regime\textquotedblright\ in \cite%
{me-arx,jpcs} and \textquotedblleft blue-sideband
transition\textquotedblright\ in \cite{bea}, recalling that the Anti
Jaynes-Cummings (AJC) Hamiltonian is $\hat{H}_{AJC}\propto \hat{a}\hat{\sigma%
}_{-}+\hat{a}^{\dagger }\hat{\sigma}_{+}$. For $N\gg 1$ this behavior turns
into the \textquotedblleft mixed behavior\textquotedblright\ described
approximately by the effective Hamiltonian $\hat{H}_{eff}\propto \hat{A}\hat{%
B}+\hat{A}^{\dagger }\hat{B}^{\dagger }$, as follows from equation (\ref%
{pari}). Moreover, for the external one-photon pumping with frequency $\eta
_{M}^{\left( p\right) }=\Omega _{0}-\left( \delta _{-}-\delta _{+}\right)
\left( 2M+1\right) $ (neglecting the nonlinearity $\alpha $) we obtain%
\begin{eqnarray*}
\dot{b}_{m,\mathcal{D}} &=&i\frac{g_{0}\varepsilon _{d}^{(p)}}{2\left\vert
\Delta _{-}\right\vert }e^{-2it\left( \delta _{-}-\delta _{+}\right) \left(
M-m\right) }b_{m+1,\mathcal{-D}}~ \\
\dot{b}_{m+1,\mathcal{-D}} &=&i\frac{g_{0}\varepsilon _{d}^{(p)\ast }}{%
2\left\vert \Delta _{-}\right\vert }e^{2it\left( \delta _{-}-\delta
_{+}\right) \left( M-m\right) }b_{m,\mathcal{D}}
\end{eqnarray*}%
under the approximations $\varepsilon _{d}\sqrt{m}\ll |\Delta _{-}|$ and $O%
(\nu _{m,\mathcal{T}}^{(3)})\sim \varepsilon _{d}^{2}/|\Delta _{-}|$. So for
$|\delta _{-}|\gg g_{0}\varepsilon _{d}/|\Delta _{-}|$ one couples only the
amplitudes $b_{M,\mathcal{D}}\leftrightarrow b_{M+1,\mathcal{-D}}$,
corresponding to the selective excitation of the atom $|g,M\rangle
\leftrightarrow |e,M\rangle $ conditioned on the presence of $M$ photons in
the cavity field. From the first line of equation (\ref{b}) we see that in
the dispersive regime one can also couple the amplitudes $b_{m,\mathcal{D}}$
and $b_{m,-\mathcal{D}}$, or roughly the states with the same number of
excitations $|g,M\rangle \leftrightarrow |e,M-1\rangle $, by employing the
modulation frequency $\approx \Delta _{-}$. This behavior was called
\textquotedblleft JC regime\textquotedblright\ in \cite{me-arx,jpcs} and
\textquotedblleft red-sideband transition\textquotedblright\ in \cite{bea}.
The generation of a single photon from vacuum and the transfer of
populations between the cavity field and the atom using the red- and
blue-sideband transitions was studied in details in \cite{jpcs,1}.

In the resonant regime, $\Delta _{-}=0$, we can couple the dressed states $%
|\varphi _{m,\mathcal{T}}\rangle \leftrightarrow |\varphi _{m+2,\mathcal{S}%
}\rangle $, where $|\varphi _{m,\pm }\rangle =\left( |g,m\rangle \pm
|e,m-1\rangle \right) /\sqrt{2}$, for any values of $m$, $\mathcal{T}$ and $%
\mathcal{S}$ by the modulation frequencies $\eta ^{(r)}=\bar{\lambda}_{m+2,%
\mathcal{S}}-\bar{\lambda}_{m,\mathcal{T}}$. The corresponding coupling
rates are of the same order of magnitude for any $\mathcal{T}$ and $\mathcal{%
S}$ and are given in \ref{lemmy}. Besides, the states $|\varphi _{m,\mathcal{%
T}}\rangle \leftrightarrow |\varphi _{m+1,\mathcal{S}}\rangle $ can be
coupled by the classical pumping with frequency $\eta ^{(p)}=\bar{\lambda}%
_{m+1,\mathcal{S}}-\bar{\lambda}_{m,\mathcal{T}}$. Therefore combining the
temporal modulation of the system parameters with the external one-photon
pumping one can create arbitrary superpositions of dressed states with a
high degree of control. Moreover, one could apply several resonant
modulation frequencies at once to study the dynamics under the multi-tone
modulation, when many dressed states are coupled simultaneously with
controllable rates.

\section{Conclusions}

\label{conclusions}

We showed analytically that Cavity dynamical Casimir effect is contained implicitly in
the most basic form of the light--matter interaction -- the dipole
interaction between a single atom and a cavity field mode under external
modulation of the atomic parameters. This phenomenon is intrinsically
nonlinear due to the nonharmonic energy spectrum of the atom--field system,
so the number of photons created from vacuum is limited and the resulting
field state can be quite different from the squeezed vacuum state. The atom
becomes entangled with the field and the average photon number exhibits
collapse-revival behavior as function of time, very sensitive to small
shifts in the modulation frequency. The effect persists when the number of
noninteracting atoms $N$ is increased, and for $N\gg 1$ our approach behaves
as a toy model for a oscillating dielectric slab inside a stationary cavity.
The precise knowledge of the atomic parameters is not required to achieve
DCE, since the photon generation occurs for the modulation frequency in the
vicinity of $2\omega _{0}$. Moreover, for additional external classical
pumping the photon production via DCE can be substantially enhanced for
appropriately chosen phase of the pump.

If the atomic parameters are known and controllable \emph{in situ} our model
describes the nonstationary circuit QED architecture (for $N\sim 1$) or cold
atomic clouds (for $N\gg 1$). In this case we can employ other modulation
frequencies to realize new effective regimes of light--matter interaction. In
the dispersive regime these frequencies and associated effects are
summarized in table \ref{table1}. In a new effect, that we called
\textquotedblleft Anti-DCE\textquotedblright ,  the modulation of atomic
parameters can lead to coherent annihilation of three photons accompanied by
the creation of one atomic excitation; however, the associated transition
rate is very small so this behavior hardly can be implemented
experimentally. Besides, we found atomic analogs of the DCE and Anti-DCE
behaviors, when the photonic and the collective atomic operators are
interchanged. Finally, we demonstrated that entangled states (dressed
atom--field states) can be generated in a straightforward manner in
nonstationary circuit QED.

\begin{table}[t]
\caption{Abbreviation of the effects with atomic clouds in the dispersive
regime. The asterisk marks the effects that persist for a single qubit. $%
\protect\eta ^{(j)}$ stands for the approximate modulation frequency. }
\label{table1}
\begin{tabular}{@{}l|l|l} \hline
Abbreviation & $\eta ^{(j)}$ & Main effect \\ \hline
DCE (*) & $2\omega _{0}$ & Generation of pairs of photons \\
IDCE & $2\Omega _{0}$ & Generation of pairs of atomic excitations \\
Mixed (*) & $\omega _{0}+\Omega _{0}$ & Equal generation of photons and
atomic excitations \\
Anti-DCE (*) & $3\omega _{0}-\Omega _{0}$ & Annihilation of three photons \\
Anti-IDCE & $3\Omega _{0}-\omega _{0}$ & Annihilation of three atomic
excitations \\ \hline
\end{tabular}
\end{table}

\appendix

\section{Analytical results for $N\gg 1$}

\label{an}

Following the method described in \cite{JPA} we write the solution for the
annihilation operators $\hat{a}$ and $\hat{b}$ in the Heisenberg picture as%
\begin{equation}
\hat{a}=\frac{e^{-it\Delta _{+}/2}}{\beta }\left[ (\beta _{+}\hat{a}_{h}+%
\tilde{g}_{0}\hat{b}_{h})e^{-it\beta /2}+(\beta _{-}\hat{a}_{h}-\tilde{g}_{0}%
\hat{b}_{h})e^{it\beta /2}\right]  \label{an1}
\end{equation}%
\begin{equation}
\hat{b}=\frac{e^{-it\Delta _{+}/2}}{\beta }\left[ (\beta _{-}\hat{b}_{h}+%
\tilde{g}_{0}\hat{a}_{h})e^{-it\beta /2}+(\beta _{+}\hat{b}_{h}-\tilde{g}_{0}%
\hat{a}_{h})e^{it\beta /2}\right] ~,  \label{an3}
\end{equation}%
where we defined time-independent parameters%
\begin{equation}
\beta \equiv \sqrt{\Delta _{-}^{2}+4\tilde{g}_{0}^{2}}~,~\Delta _{\pm
}\equiv \omega _{0}\pm \Omega _{0}~,~\beta _{\pm }\equiv \frac{\beta \pm
\Delta _{-}}{2}~,~\varepsilon _{\pm }^{(j)}\equiv \varepsilon _{\omega
}^{(j)}\pm \varepsilon _{\Omega }^{(j)}  \label{parameters}
\end{equation}%
so that $\Delta _{-}$ stands for the bare atom--cavity detuning.

The auxiliary annihilation operators $\hat{a}_{h}$ and $\hat{b}_{h}$ satisfy
the bosonic commutation relations $[\hat{a}_{h},\hat{a}_{h}^{\dagger }]=1$, $%
[\hat{b}_{h},\hat{b}_{h}^{\dagger }]=1$, $[\hat{a}_{h},\hat{b}_{h}]=[\hat{a}%
_{h},\hat{b}_{h}^{\dagger }]=0$. Under the approximations%
\begin{equation}
\frac{\tilde{\varepsilon}_{g}}{\tilde{g}_{0}}\lesssim 1~;~\frac{\varepsilon
_{\omega }}{\omega _{0}},\frac{\varepsilon _{\Omega }}{\omega _{0}},\frac{%
\Delta _{-}}{\omega _{0}},\frac{\varepsilon _{d}}{\omega _{0}},\frac{\langle
\hat{b}^{\dagger }\hat{b}\rangle }{N},\frac{\varepsilon _{\omega ,\Omega
,g}^{(j\prime \prime )}}{\beta }\sqrt{\mathbf{E}},\frac{\tilde{g}_{0}}{%
\omega _{0}}\sqrt{\mathbf{E}},\frac{\chi _{0}}{\omega _{0}}\sqrt{\mathbf{E}},%
\frac{\varepsilon _{\chi }}{\omega _{0}}\sqrt{\mathbf{E}}\ll 1~
\label{appro-Heis}
\end{equation}%
they read%
\begin{eqnarray}
\hat{a}_{h} &=&e^{it\tilde{\delta}_{+}}\left( \hat{A}e^{it\left( \delta
_{\chi }-\tilde{\delta}_{s}\right) }+i\mathcal{F}_{AB}\hat{B}e^{it\tilde{%
\delta}_{s}}\right) e^{i\mathcal{F}_{A}}~  \nonumber \\
\hat{b}_{h} &=&e^{it\tilde{\delta}_{+}}\left( \hat{B}e^{it\tilde{\delta}%
_{s}}+i\mathcal{F}_{AB}^{\ast }\hat{A}e^{it\left( \delta _{\chi }-\tilde{%
\delta}_{s}\right) }\right) e^{i\mathcal{F}_{B}}~.  \label{relation}
\end{eqnarray}%
Here $\mathbf{E}\equiv \langle \hat{a}^{\dagger }\hat{a}\rangle +\langle
\hat{b}^{\dagger }b\rangle $ denotes the total number of excitations in the
atoms--field system and we defined small \textquotedblleft intrinsic
frequency shifts\textquotedblright
\begin{equation}
\tilde{\delta}_{\pm }\equiv \frac{\tilde{g}_{0}^{2}}{\Delta _{\pm }}%
~,~\delta _{\chi }=\frac{4\chi _{0}^{2}}{\Delta _{+}}~,~\tilde{\delta}%
_{s}\equiv \sum\nolimits_{j}^{\prime \prime }\frac{\tilde{g}_{0}}{\beta ^{2}}%
{\rm Im}\left( \tilde{g}_{0}\varepsilon _{-}^{(j)}-\Delta _{-}\tilde{\varepsilon%
}_{g}^{(j)}\right) ~.  \label{shifts}
\end{equation}%
$\tilde{\delta}_{-}$ is the standard dispersive shift, $\tilde{\delta}_{+}$
is the collective Bloch-Siegert shift, $\delta _{\chi }$ is the shift due to
the term $i\chi _{0}(\hat{a}^{\dagger 2}-\hat{a}^{2})$ and $\tilde{\delta}%
_{s}$ is the shift due to a possible modulation of the system parameters
with a low modulation frequency $\eta ^{(j\prime \prime )}\approx \beta $.

The independent annihilation operators $\hat{A}$ and $\hat{B}$, that also
satisfy the bosonic commutation relations, are defined implicitly in terms
of the small time-dependent functions $|\mathcal{F}_{A}|,|\mathcal{F}_{B}|,|%
\mathcal{F}_{AB}|\ll 1$%
\begin{eqnarray}
\mathcal{F}_{A} &=&\sum\nolimits_{j}^{\prime }\frac{1}{2\beta ^{2}}\left[
(\varepsilon _{\omega }^{(j)}\beta ^{2}-2\varepsilon _{-}^{(j)}\tilde{g}%
_{0}^{2}+2\tilde{\varepsilon}_{g}^{(j)}\tilde{g}_{0}\Delta _{-})\frac{%
e^{it\eta ^{(j)}}-1}{\eta ^{(j)}}\right. \\
&&\left. +(\varepsilon _{-}^{(j)}\tilde{g}_{0}^{2}-\tilde{\varepsilon}%
_{g}^{(j)}\tilde{g}_{0}\Delta _{-})\frac{e^{it(\eta ^{(j)}+\beta )}-1}{\eta
^{(j)}+\beta }+(\varepsilon _{-}^{(j)}\tilde{g}_{0}^{2}-\tilde{\varepsilon}%
_{g}^{(j)}\tilde{g}_{0}\Delta _{-})\frac{e^{it(\eta ^{(j)}-\beta )}-1}{\eta
^{(j)}-\beta }+c.c.\right]  \nonumber
\end{eqnarray}%
\begin{eqnarray}
\mathcal{F}_{B} &=&\sum\nolimits_{j}^{\prime }\frac{1}{2\beta ^{2}}\left[
(\varepsilon _{\Omega }^{(j)}\beta ^{2}+2\varepsilon _{-}^{(j)}\tilde{g}%
_{0}^{2}-2\tilde{\varepsilon}_{g}^{(j)}\Delta _{-}\tilde{g}_{0})\frac{%
e^{it\eta ^{(j)}}-1}{\eta ^{(j)}}\right. \\
 &&\left. -(\varepsilon _{-}^{(j)}\tilde{g}_{0}^{2}-\tilde{\varepsilon}%
_{g}^{(j)}\tilde{g}_{0}\Delta _{-})\frac{e^{it(\eta ^{(j)}+\beta )}-1}{\eta
^{(j)}+\beta }-(\varepsilon _{-}^{(j)}\tilde{g}_{0}^{2}-\tilde{\varepsilon}%
_{g}^{(j)}\tilde{g}_{0}\Delta _{-})\frac{e^{it(\eta ^{(j)}-\beta )}-1}{\eta
^{(j)}-\beta }+c.c.\right]  \nonumber
\end{eqnarray}%
\begin{eqnarray}
\mathcal{F}_{AB} &=&\sum\nolimits_{j}^{\prime }\frac{1}{2\beta ^{2}}\left[
(\varepsilon _{-}^{(j)}\tilde{g}_{0}\Delta _{-}+4\tilde{\varepsilon}%
_{g}^{(j)}\tilde{g}_{0}^{2})\frac{e^{it\eta ^{(j)}}-1}{\eta ^{(j)}}%
+(\varepsilon _{-}^{(j)\ast }\tilde{g}_{0}\Delta _{-}+4\tilde{\varepsilon}%
_{g}^{(j)\ast }\tilde{g}_{0}^{2})\frac{e^{-it\eta ^{(j)}}-1}{\eta ^{(j)}}%
\right.  \nonumber \\
 &&+(\varepsilon _{-}^{(j)}\tilde{g}_{0}-\tilde{\varepsilon}%
_{g}^{(j)}\Delta _{-})\left( \beta _{-}\frac{e^{it(\eta ^{(j)}-\beta )}-1}{%
\eta ^{(j)}-\beta }-\beta _{+}\frac{e^{it(\eta ^{(j)}+\beta )}-1}{\eta
^{(j)}+\beta }\right) \\
 &&+\left. (\varepsilon _{-}^{(j)\ast }\tilde{g}_{0}-\tilde{\varepsilon}%
_{g}^{(j)\ast }\Delta _{-})\left( \beta _{-}\frac{e^{-it(\eta ^{(j)}+\beta
)}-1}{\eta ^{(j)}+\beta }-\beta _{+}\frac{e^{-it(\eta ^{(j)}-\beta )}-1}{%
\eta ^{(j)}-\beta }\right) \right] ~.  \nonumber
\end{eqnarray}%
The time evolution of the operators $\hat{A}$ and $\hat{B}$ is governed by
the Heisenberg equation of motion $id\hat{O}/dt=[\hat{O},\hat{H}_{eff}]$,
where $\hat{O}=\hat{A},\hat{B}$ and the effective Hamiltonian can be written
as%
\begin{equation}
\hat{H}_{eff}=\hat{H}_{G}+\hat{H}_{NG}(\tilde{g}_{0})+\sum_{X=g,\omega
,\Omega }\hat{H}_{NG}(\varepsilon _{X})~.  \label{effective}
\end{equation}%
$\hat{H}_{G}$ denotes the Gaussian part containing linear and quadratic
combinations of $\hat{A}$ and $\hat{B}$. $\hat{H}_{NG}(\tilde{g}_{0})$ and $%
\hat{H}_{NG}(\varepsilon _{X})$ denote the non-Gaussian parts, of the fourth
order in operators $\hat{A}$ and $\hat{B}$, proportional to $\tilde{g}_{0}$
and $\varepsilon _{X}$, respectively. For simplicity we shall consider only
the term $\hat{H}_{NG}(\tilde{\varepsilon}_{g})$ in the last term of
equation (\ref{effective}), since the resulting general expressions are too
long to write out explicitly.

Eliminating the rapidly oscillating terms via the Rotating Wave
approximation (RWA) \cite{JPA} we obtain for the Gaussian part%
\begin{eqnarray}
\hat{H}_{G} &=&\frac{i}{2\beta ^{2}}\sum\nolimits_{j}^{\prime \prime }%
\left[ \beta _{+}\left( \varepsilon _{-}^{(j)}\tilde{g}_{0}-\tilde{%
\varepsilon}_{g}^{(j)}\Delta _{-}\right) e^{-it\left( \beta +2\tilde{\delta}%
_{s}-\delta _{\chi }-\eta ^{(j)}\right) }\right.  \nonumber \\
 &&\left. +\beta _{-}\left( \varepsilon _{-}^{(j)\ast }\tilde{g}_{0}-%
\tilde{\varepsilon}_{g}^{(j)\ast }\Delta _{-}\right) e^{it\left( \beta -2%
\tilde{\delta}_{s}+\delta _{\chi }-\eta ^{(j)}\right) }\hat{A}\hat{B}%
^{\dagger }\right]  \nonumber \\
 &&+\sum\nolimits_{j}^{\prime }e^{-it\left( \Delta _{+}-2\tilde{\delta}%
_{+}-\eta ^{(j)}\right) }\left\{ \frac{1}{2}e^{2it\left( \delta _{\chi }-%
\tilde{\delta}_{s}\right) }\left[ \tilde{g}_{0}\Theta _{0}^{(j)}+\beta
_{+}\Theta _{+}^{(j)}e^{-it\beta }+\beta _{-}\Theta _{-}^{(j)}e^{it\beta }%
\right] \hat{A}^{2}\right.  \nonumber \\
 &&+\frac{1}{2}e^{2it\tilde{\delta}_{s}}\left[ -\tilde{g}_{0}\Theta
_{0}^{(j)}+\beta _{-}\Theta _{+}^{(j)}e^{-it\beta }+\beta _{+}\Theta
_{-}^{(j)}e^{it\beta }\right] \hat{B}^{2}  \nonumber \\
 &&\left. +e^{it\delta _{\chi }}\left[ -\frac{1}{2}\Delta _{-}\Theta
_{0}^{(j)}+\tilde{g}_{0}\Theta _{+}^{(j)}e^{-it\beta }-\tilde{g}_{0}\Theta
_{-}^{(j)}e^{it\beta }\right] \hat{A}\hat{B}\right\}  \nonumber \\
 &&-\frac{1}{2\beta }\sum\nolimits_{j}^{\prime }e^{-it\left( \Delta
_{+}/2-\tilde{\delta}_{+}-\eta ^{(j)}\right) }\left\{ \varepsilon
_{d}^{(j)}e^{it\left( \delta _{\chi }-\tilde{\delta}_{s}\right) }\left[
\beta _{-}e^{it\beta /2}+\beta _{+}e^{-it\beta /2}\right] \hat{A}\right.
\nonumber \\
 &&\left. +\varepsilon _{d}^{(j)}e^{-it\tilde{\delta}_{s}}\left[ \tilde{g}%
_{0}e^{-it\beta /2}-\tilde{g}_{0}e^{it\beta /2}\right] \hat{B}\right\} +h.c.
\label{HG}
\end{eqnarray}%
The time-independent coefficients are%
\begin{equation}
\Theta _{0}^{(j)}=-iW_{0}^{(j)}-2\chi _{0}V_{0}^{(j)}~,~\Theta _{\pm
}^{(j)}=-iW_{\pm }^{(j)}-\chi _{0}V_{\pm }^{(j)}
\end{equation}%
\begin{equation}
W_{0}^{(j)}=\frac{1}{\beta ^{2}}\left[ \Delta _{-}\tilde{g}_{0}\left( \frac{%
\varepsilon _{+}^{(j)}}{\eta ^{(j)}}-\frac{\tilde{\varepsilon}_{g}^{(j)}}{%
\tilde{g}_{0}}\right) +\frac{4\tilde{g}_{0}^{2}}{\eta ^{(j)}-\beta ^{2}}%
(\varepsilon _{-}^{(j)}\tilde{g}_{0}-\tilde{\varepsilon}_{g}^{(j)}\Delta
_{-})\right]
\end{equation}%
\begin{equation}
V_{0}^{(j)}=\frac{1}{\beta ^{2}}\left[ \frac{\varepsilon _{\chi }^{(j)}%
\tilde{g}_{0}}{\chi _{0}}-\left( \frac{\tilde{g}_{0}\varepsilon _{+}^{(j)}}{%
\eta ^{(j)}}+\frac{\eta ^{(j)}-\Delta _{-}}{\eta ^{(j)}-\beta ^{2}}%
(\varepsilon _{-}^{(j)}\tilde{g}_{0}-\tilde{\varepsilon}_{g}^{(j)}\Delta
_{-})\right) \right]
\end{equation}%
\begin{equation}
W_{\pm }^{(j)}=\frac{1}{\beta ^{2}}\left[ (\varepsilon _{-}^{(j)}\tilde{g}%
_{0}-\tilde{\varepsilon}_{g}^{(j)}\Delta _{-})\frac{\tilde{g}_{0}\Delta _{-}%
}{\beta \left( \eta ^{(j)}\mp \beta \right) }\mp (\varepsilon
_{+}^{(j)}\beta _{\mp }\pm 2\tilde{\varepsilon}_{g}^{(j)}\tilde{g}_{0}\pm
\varepsilon _{\omega }^{(j)}\Delta _{-})\frac{2\tilde{g}_{0}^{2}}{\beta \eta
^{(j)}}\pm \tilde{\varepsilon}_{g}^{(j)}\tilde{g}_{0}\right]
\end{equation}%
\begin{equation}
V_{\pm }^{(j)}=\frac{1}{\beta ^{2}}\left[ \frac{\varepsilon _{\chi
}^{(j)}\beta _{\pm }}{\chi _{0}}-\frac{2}{\beta }\left( (\varepsilon
_{\omega }^{(j)}\beta _{\pm }+\varepsilon _{\Omega }^{(j)}\beta _{\mp }\pm 2%
\tilde{\varepsilon}_{g}^{(j)}\tilde{g}_{0})\frac{\beta _{\pm }}{\eta ^{(j)}}%
-(\tilde{\varepsilon}_{g}^{(j)}\Delta _{-}-\varepsilon _{-}^{(j)}\tilde{g}%
_{0})\frac{\tilde{g}_{0}}{\eta ^{(j)}\mp \beta }\right) \right] .
\end{equation}

Due to elimination of the rapidly oscillating terms one introduces intrinsic
uncertainty $\Delta \nu $ in the arguments of the exponential functions in
equation (\ref{HG}) of the order%
\begin{equation}
O(\Delta \nu )\sim \left( \frac{\tilde{g}_{0}}{\beta }\right) ^{2}\frac{%
\varepsilon _{\omega }^{2}}{\omega _{0}},\left( \frac{\tilde{g}_{0}}{\beta }%
\right) ^{2}\frac{\varepsilon _{\Omega }^{2}}{\omega _{0}},~\frac{\tilde{%
\varepsilon}_{g}^{2}}{\omega _{0}},\frac{\tilde{g}_{0}^{2}}{\omega _{0}}%
\left( \frac{\beta }{\omega _{0}}\right) ^{2},\frac{\varepsilon _{\chi }^{2}%
}{\omega _{0}},\frac{\chi _{0}^{2}}{\omega _{0}}\left( \frac{\beta }{\omega
_{0}}\right) ^{2}.
\end{equation}%
We call these contributions \textquotedblleft Systematic-error frequency
shifts\textquotedblright\ (SEFS), since they appear due to systematic
simplification of the differential equations for $\hat{a}_{h}$ and $\hat{b}%
_{h}$. In practice SEFS slightly alter the resonant modulation frequencies $%
\eta ^{(j)}$ that give rise to nontrivial behavior, so ultimately they must
be found experimentally or numerically.

Under the additional approximations%
\begin{equation}
\frac{\mathbf{E}\sqrt{\mathbf{E}}}{N}\ll 1\mbox{ (Resonant regime: }\Delta
_{-}=0\mbox{)}
\end{equation}%
\begin{equation}
\frac{\tilde{g}_{0}}{\Delta _{-}}\frac{\sqrt{\mathbf{E}}}{N}\left\{ \langle
\hat{b}^{\dagger }\hat{b}\rangle ,\sqrt{\langle \hat{a}^{\dagger }\hat{a}%
\rangle \langle \hat{b}^{\dagger }\hat{b}\rangle },\frac{\tilde{g}_{0}}{%
\Delta _{-}}\langle \hat{a}^{\dagger }\hat{a}\rangle \right\} \ll 1%
\mbox{
(Dispersive regime: }\left\vert \Delta _{-}\right\vert /2\gg \tilde{g}_{0}%
\mbox{)}  \label{disp}
\end{equation}%
one obtains for the first non-Gaussian term in equation (\ref{effective})%
\begin{eqnarray}
\hat{H}_{NG}(\tilde{g}_{0}) &=&-\frac{1}{2N}\frac{\tilde{g}_{0}^{2}}{%
\beta ^{4}}\left[ 3\tilde{g}_{0}^{2}\Delta _{-}\hat{A}^{\dagger 2}\hat{A}%
^{2}+2\Delta _{-}\left( \Delta _{-}^{2}-2\tilde{g}_{0}^{2}\right) \hat{A}%
^{\dagger }\hat{A}\hat{B}^{\dagger }\hat{B}\right.  \label{HNG} \\
 &&+8\tilde{g}_{0}\left( \Delta _{-}^{2}+\tilde{g}_{0}^{2}\right) \hat{A}%
^{\dagger }\hat{B}^{\dagger }\hat{B}^{2}e^{-it\left( \delta _{\chi }-2\tilde{%
\delta}_{s}\right) }-6\tilde{g}_{0}^{2}\Delta _{-}\hat{A}^{2}\hat{B}%
^{\dagger 2}e^{2it\left( \delta _{\chi }-2\tilde{\delta}_{s}\right) }  \nonumber
\\
 &&\left. -4\tilde{g}_{0}\left( \Delta _{-}^{2}-2\tilde{g}_{0}^{2}\right)
\hat{A}^{\dagger 2}\hat{A}\hat{B}e^{-it\left( \delta _{\chi }-2\tilde{\delta}%
_{s}\right) }-\Delta _{-}\left( \Delta _{-}^{2}+\tilde{g}_{0}^{2}\right)
\hat{B}^{\dagger 2}\hat{B}^{2}\right] +h.c.  \nonumber
\end{eqnarray}%
The simplified expression for the non-Gaussian term $\hat{H}_{NG}(\tilde{%
\varepsilon}_{g})$ strongly depends on the modulation frequency $\eta ^{(j)}$%
, as can be seen from equation (\ref{NHN}), so we do not write it explicitly
due to its length. In \ref{dispersive} we shall give the approximate results
for $\hat{H}_{NG}(\tilde{\varepsilon}_{g})$ in the dispersive regime for
high modulation frequencies $\eta ^{(j)}\sim 2\omega _{0}$.

\subsection{Simplified formulae in the resonant regime}

\label{resonant}

For $\Delta _{-}=0$ we obtain the simplified expressions%
\begin{eqnarray}
\hat{a} &\simeq &e^{-it\left( \omega _{0}-\tilde{\delta}_{+}\right) }\left[
\hat{A}e^{it\left( \delta _{\chi }-\tilde{\delta}_{s}\right) }\cos \tilde{g}%
_{0}t-i\hat{B}e^{it\tilde{\delta}_{s}}\sin \tilde{g}_{0}t\right] \\
\hat{b} &\simeq &e^{-it\left( \omega _{0}-\tilde{\delta}_{+}\right) }\left[
\hat{B}e^{it\tilde{\delta}_{s}}\cos \tilde{g}_{0}t-i\hat{A}e^{it\left(
\delta _{\chi }-\tilde{\delta}_{s}\right) }\sin \tilde{g}_{0}t\right] .
\end{eqnarray}%
The Hamiltonians are
\begin{eqnarray}
\hat{H}_{G} &\simeq& i\sum\nolimits_{j}^{\prime \prime }\frac{1}{8}\left(
\varepsilon _{-}^{(j)}e^{-it\left( 2\tilde{g}_{0}+2\tilde{\delta}_{s}-\delta
_{\chi }-\eta ^{(j)}\right) }+\varepsilon _{-}^{(j)\ast }e^{it\left( 2\tilde{%
g}_{0}-2\tilde{\delta}_{s}+\delta _{\chi }-\eta ^{(j)}\right) }\right) \hat{A%
}\hat{B}^{\dagger }  \nonumber \\
&&-\frac{1}{4}\sum\nolimits_{j}^{\prime }e^{-it\left( \omega _{0}-\tilde{%
\delta}_{+}-\eta ^{(j)}\right) }\left\{ \varepsilon _{d}^{(j)}e^{it\left(
\delta _{\chi }-\tilde{\delta}_{s}\right) }\left( e^{it\tilde{g}_{0}}+e^{-it%
\tilde{g}_{0}}\right) \hat{A}\right.  \nonumber \\
&&+\left. \varepsilon _{d}^{(j)}e^{it\tilde{\delta}_{s}}\left( e^{-it\tilde{g%
}_{0}}-e^{it\tilde{g}_{0}}\right) \hat{B}\right\}  \nonumber \\
&&+\tilde{g}_{0}\sum\nolimits_{j}^{\prime }e^{-it\left( 2\omega _{0}-2\tilde{%
\delta}_{+}-\eta ^{(j)}\right) }\left\{ e^{it\delta _{\chi }}\left( \Theta
_{+}^{(j)}e^{-it2\tilde{g}_{0}}-\Theta _{-}^{(j)}e^{it2\tilde{g}_{0}}\right)
\hat{A}\hat{B}\right.  \nonumber \\
&&+\frac{1}{2}e^{2it\left( \delta _{\chi }-\tilde{\delta}_{s}\right) }\left(
\Theta _{0}^{(j)}+\Theta _{+}^{(j)}e^{-it2\tilde{g}_{0}}+\Theta
_{-}^{(j)}e^{it2\tilde{g}_{0}}\right) \hat{A}^{2}  \nonumber \\
&&\left. +\frac{1}{2}e^{2it\tilde{\delta}_{s}}\left( -\Theta
_{0}^{(j)}+\Theta _{+}^{(j)}e^{-it2\tilde{g}_{0}}+\Theta _{-}^{(j)}e^{it2%
\tilde{g}_{0}}\right) \hat{B}^{2}\right\} +h.c.
\end{eqnarray}%
\begin{equation}
\hat{H}_{NG}(\tilde{g}_{0})=-\frac{\tilde{g}_{0}}{8N}\left( \hat{A}^{\dagger
}\hat{A}+\hat{B}^{\dagger }\hat{B}-1\right) \left[ \hat{A}\hat{B}^{\dagger
}e^{it\left( \delta _{\chi }-2\tilde{\delta}_{s}\right) }+\hat{A}^{\dagger }%
\hat{B}e^{-it\left( \delta _{\chi }-2\tilde{\delta}_{s}\right) }\right] ~.
\end{equation}%
The time-independent coefficients become%
\begin{eqnarray}
\Theta _{0}^{(j)} &=&\frac{1}{2\tilde{g}_{0}}\left( \frac{2\chi _{0}}{\eta
^{(j)}}\varepsilon _{\omega }^{(j)}-\varepsilon _{\chi }^{(j)}\right) -i%
\frac{\tilde{g}_{0}}{(\eta ^{(j)})^{2}}\varepsilon _{-}^{(j)}  \nonumber \\
\Theta _{\pm }^{(j)} &=&\frac{1}{4}\left[ \frac{1}{\eta ^{(j)}}\left( \frac{%
2\chi _{0}}{\tilde{g}_{0}}\varepsilon _{\omega }^{(j)}\pm i\varepsilon
_{+}^{(j)}\right) -\frac{\varepsilon _{\chi }^{(j)}\pm i\tilde{\varepsilon}%
_{g}^{(j)}}{\tilde{g}_{0}}\right] ~.  \label{guest}
\end{eqnarray}

\subsection{Simplified formulae in the dispersive regime}

\label{dispersive}

In the dispersive regime, $|\Delta _{-}|/2\gg \tilde{g}_{0}$, we have $\beta
\approx |\Delta _{-}|+2|\tilde{\delta}_{-}|$, where $\tilde{\delta}_{-}=%
\tilde{g}_{0}^{2}/\Delta _{-}$ is the collective dispersive shift. The
operators read approximately%
\begin{eqnarray}
\hat{a} &\simeq &e^{-it\left( \omega _{0}+\tilde{\delta}_{s}-\tilde{\delta}%
_{+}-\delta _{\chi }\right) }\left[ \hat{A}-i\hat{B}\frac{2\tilde{g}_{0}}{%
\Delta _{-}}e^{it\left( \Delta _{-}/2+2\tilde{\delta}_{s}-\delta _{\chi
}\right) }\sin \left( \Delta _{-}t/2\right) \right]  \nonumber \\
\hat{b} &\simeq &e^{-it\left( \Omega _{0}-\tilde{\delta}_{s}-\tilde{\delta}%
_{+}\right) }\left[ \hat{B}-i\hat{A}\frac{2\tilde{g}_{0}}{\Delta _{-}}%
e^{-it\left( \Delta _{-}/2+2\tilde{\delta}_{s}-\delta _{\chi }\right) }\sin
\left( \Delta _{-}t/2\right) \right] ~.
\end{eqnarray}%
Introducing the \textquotedblleft detuning symbol\textquotedblright
\begin{equation}
\mathcal{D}\equiv \frac{\Delta _{-}}{\left\vert \Delta _{-}\right\vert }=\pm
~  \label{dsym}
\end{equation}%
the Hamiltonians can be written as%
\begin{eqnarray}
\hat{H}_{G} &\simeq& \frac{i}{2}\mathcal{D}\sum\nolimits_{j}^{\prime
\prime }\left[ \left( \varepsilon _{-}^{(j)}\frac{\tilde{g}_{0}}{\Delta _{-}}%
-\tilde{\varepsilon}_{g}^{(j)}\right) _{\mathcal{D}}e^{-it\left( \Delta
_{-}+2\tilde{\delta}_{-}+2\tilde{\delta}_{s}-\delta _{\chi }-\mathcal{D}\eta
^{(j)}\right) }\right.  \nonumber \\
&&\left. +\frac{\tilde{g}_{0}^{2}}{\Delta _{-}^{2}}\left( \varepsilon
_{-}^{(j)}\frac{\tilde{g}_{0}}{\Delta _{-}}-\tilde{\varepsilon}%
_{g}^{(j)}\right) _{-\mathcal{D}}e^{it\left( \Delta _{-}+2\tilde{\delta}%
_{-}-2\tilde{\delta}_{s}+\delta _{\chi }-\mathcal{D}\eta ^{(j)}\right) }%
\right] \hat{A}\hat{B}^{\dagger }  \nonumber \\
&&-\frac{1}{2}\sum\nolimits_{j}^{\prime }e^{-it\left( \Delta _{+}/2-\tilde{%
\delta}_{+}-\eta ^{(j)}\right) }\left[ \varepsilon _{d}^{(j)}e^{it\left(
\delta _{\chi }-\tilde{\delta}_{s}\right) }\left( e^{-it\left( \Delta _{-}+2%
\tilde{\delta}_{-}\right) /2}+\frac{\tilde{g}_{0}^{2}}{\Delta _{-}^{2}}%
e^{it\left( \Delta _{-}+2\tilde{\delta}_{-}\right) /2}\right) \hat{A}\right.
\nonumber \\
&&+\left. \frac{\tilde{g}_{0}}{\Delta _{-}}\varepsilon _{d}^{(j)}e^{it\tilde{%
\delta}_{s}}\left( e^{-it\left( \Delta _{-}+2\tilde{\delta}_{-}\right)
/2}-e^{it\left( \Delta _{-}+2\tilde{\delta}_{-}\right) /2}\right) \hat{B}%
\right] +\sum\nolimits_{j}^{\prime }e^{-it\left( \Delta _{+}-2\tilde{\delta}%
_{+}-\eta ^{(j)}\right) }  \nonumber \\
&&\times \left\{ \frac{1}{2}e^{2it\left( \delta _{\chi }-\tilde{\delta}%
_{s}\right) }\left[ \tilde{g}_{0}\Theta _{0}^{(j)}+|\Delta _{-}|\Theta _{%
\mathcal{D}}^{(j)}e^{-it\left( \Delta _{-}+2\tilde{\delta}_{-}\right) }+|%
\tilde{\delta}_{-}|\Theta _{-\mathcal{D}}^{(j)}e^{it\left( \Delta _{-}+2%
\tilde{\delta}_{-}\right) }\right] \hat{A}^{2}\right.  \nonumber \\
&&+\mathcal{D}e^{it\delta _{\chi }}\left[ -\frac{1}{2}|\Delta _{-}|\Theta
_{0}^{(j)}+\tilde{g}_{0}\Theta _{\mathcal{D}}^{(j)}e^{-it\left( \Delta _{-}+2%
\tilde{\delta}_{-}\right) }-\tilde{g}_{0}\Theta _{-\mathcal{D}%
}^{(j)}e^{it\left( \Delta _{-}+2\tilde{\delta}_{-}\right) }\right] \hat{A}%
\hat{B}  \label{fline} \\
&&+\left. \frac{1}{2}e^{2it\tilde{\delta}_{s}}\left[ -\tilde{g}_{0}\Theta
_{0}^{(j)}+|\tilde{\delta}_{-}|\Theta _{\mathcal{D}}^{(j)}e^{-it\left(
\Delta _{-}+2\tilde{\delta}_{-}\right) }+|\Delta _{-}|\Theta _{-\mathcal{D}%
}^{(j)}e^{it\left( \Delta _{-}+2\tilde{\delta}_{-}\right) }\right] \hat{B}%
^{2}\right\} +h.c.  \nonumber
\end{eqnarray}%
\begin{eqnarray}
\hat{H}_{NG}(\tilde{g}_{0}) \simeq \frac{\tilde{\delta}_{-}}{2N}\left[
\hat{B}^{\dagger 2}\hat{B}^{2}-2\hat{A}^{\dagger }\hat{A}\hat{B}^{\dagger }%
\hat{B}-3\frac{\tilde{g}_{0}^{2}}{\Delta _{-}^{2}}\hat{A}^{\dagger 2}\hat{A}%
^{2}-8\frac{\tilde{g}_{0}}{\Delta _{-}}\hat{A}^{\dagger }\hat{B}^{\dagger }%
\hat{B}^{2}e^{-it\left( \delta _{\chi }-2\tilde{\delta}_{s}\right) }\right.
\nonumber \\
\left. +6\frac{\tilde{g}_{0}^{2}}{\Delta _{-}^{2}}\hat{A}^{\dagger 2}\hat{B%
}^{2}e^{-2it\left( \delta _{\chi }-2\tilde{\delta}_{s}\right) }+4\frac{%
\tilde{g}_{0}}{\Delta _{-}}\hat{A}^{\dagger 2}\hat{A}\hat{B}e^{-it\left(
\delta _{\chi }-2\tilde{\delta}_{s}\right) }\right] +h.c.~~~~  \label{IDCE1}
\end{eqnarray}%
In the first line of equation (\ref{fline}) we introduced the notation $%
\left( O\right) _{+}\equiv O$ and $\left( O\right) _{-}\equiv O^{\ast }$.
The time-independent coefficients in $\hat{H}_{G}$ are%
\begin{equation}
\Theta _{0}^{(j)}\approx -\frac{\tilde{g}_{0}}{\Delta _{-}}\left[ \left( i-%
\frac{4\chi _{0}}{\Delta _{-}}\right) \frac{\varepsilon _{\omega }^{(j)}}{%
\eta ^{(j)}}+i\frac{\varepsilon _{\Omega }^{(j)}}{\eta ^{(j)}}-i\frac{\tilde{%
\varepsilon}_{g}^{(j)}}{\tilde{g}_{0}}+2\frac{\varepsilon _{\chi }^{(j)}}{%
\Delta _{-}}\right]
\end{equation}%
\begin{equation}
\Theta _{\mathcal{D}}^{(j)}\approx \frac{1}{\left\vert \Delta
_{-}\right\vert }\left[ \left( 2\chi _{0}+i\tilde{\delta}_{-}\frac{\eta
^{(j)}-2\Delta _{-}}{\eta ^{(j)}-\Delta _{-}}\right) \frac{\varepsilon
_{\omega }^{(j)}}{\eta ^{(j)}}+i\tilde{\delta}_{-}\frac{\varepsilon _{\Omega
}^{(j)}}{\eta ^{(j)}-\Delta _{-}}-i\tilde{\delta}_{-}\frac{\eta
^{(j)}-2\Delta _{-}}{\eta ^{(j)}-\Delta _{-}}\frac{\tilde{\varepsilon}%
_{g}^{(j)}}{\tilde{g}_{0}}-\varepsilon _{\chi }^{(j)}\right]
\end{equation}%
\begin{equation}
\Theta _{-\mathcal{D}}^{(j)}\approx -\mathcal{D}\frac{\tilde{g}_{0}^{2}}{%
\Delta _{-}^{2}}\left[ \left( i-\frac{2\chi _{0}}{\Delta _{-}}\right) \frac{%
\varepsilon _{\omega }^{(j)}}{\eta ^{(j)}+\Delta _{-}}+i\frac{\eta
^{(j)}+2\Delta _{-}}{\eta ^{(j)}+\Delta _{-}}\frac{\varepsilon _{\Omega
}^{(j)}}{\eta ^{(j)}}-i\frac{\eta ^{(j)}+2\Delta _{-}}{\eta ^{(j)}+\Delta
_{-}}\frac{\tilde{\varepsilon}_{g}^{(j)}}{\tilde{g}_{0}}+\frac{\varepsilon
_{\chi }^{(j)}}{\Delta _{-}}\right] ~.
\end{equation}

Neglecting the rapidly oscillating terms under the approximations (\ref{disp}%
) we can obtain particular expressions for $\hat{H}_{NG}\left( \tilde{%
\varepsilon}_{g}\right) $ for concrete modulation frequencies:

\noindent $\bullet $ for the DCE modulation frequency $\eta ^{(D)}\approx
2\omega _{0}$%
\begin{eqnarray}
\hat{H}_{NG}\left( \tilde{\varepsilon}_{g}\right) &=&i\frac{\tilde{\delta}%
_{-}}{4N}\frac{\tilde{\varepsilon}_{g}^{(D)}}{\tilde{g}_{0}}e^{-it\left[
2\left( \omega _{0}+\tilde{\delta}_{-}-\tilde{\delta}_{+}-\delta _{\chi
}\right) -\eta ^{(D)}\right] }\left[ 2\hat{A}^{2}\hat{B}^{\dagger }\hat{B}%
e^{-2it\tilde{\delta}_{s}}+4\frac{\tilde{g}_{0}}{\Delta _{-}}\hat{A}\hat{B}%
^{\dagger }\hat{B}^{2}e^{-it\delta _{\chi }}\right.  \nonumber \\
 &&-2\frac{\tilde{g}_{0}}{\Delta _{-}}\hat{A}^{\dagger }\hat{A}^{2}\hat{B}%
e^{-it\delta _{\chi }}-4\frac{\tilde{g}_{0}^{2}}{\Delta _{-}^{2}}\hat{A}%
^{\dagger }\hat{A}\hat{B}^{2}e^{-2it\left( \delta _{\chi }-\tilde{\delta}%
_{s}\right) }-2\frac{\tilde{g}_{0}}{\Delta _{-}}\hat{A}^{3}\hat{B}^{\dagger
}e^{it\left( \delta _{\chi }-4\tilde{\delta}_{s}\right) } \\
 &&\left. +3\frac{\tilde{g}_{0}^{2}}{\Delta _{-}^{2}}\hat{A}^{\dagger }%
\hat{A}^{3}e^{-2it\tilde{\delta}_{s}}-2\frac{\tilde{g}_{0}^{3}}{\Delta
_{-}^{3}}\hat{A}^{\dagger }\hat{B}^{3}e^{-it\left( 3\delta _{\chi }-4\tilde{%
\delta}_{s}\right) }+2\frac{\tilde{g}_{0}^{2}}{\Delta _{-}^{2}}\hat{B}%
^{\dagger }\hat{B}^{3}e^{-2it\left( \delta _{\chi }-\tilde{\delta}%
_{s}\right) }\right] +h.c.  \nonumber
\end{eqnarray}%
$\bullet $ for the IDCE modulation frequency $\eta ^{(I)}\approx 2\Omega
_{0} $%
\begin{eqnarray}
\hat{H}_{NG}\left( \tilde{\varepsilon}_{g}\right) &=&i\frac{\tilde{\delta}%
_{-}}{4N}\frac{\tilde{\varepsilon}_{g}^{(I)}}{\tilde{g}_{0}}e^{-it\left[
2\left( \Omega _{0}-\tilde{\delta}_{-}-\tilde{\delta}_{+}-\tilde{\delta}%
_{s}\right) -\eta ^{(I)}\right] }\left[ -\hat{B}^{\dagger }\hat{B}^{3}+\hat{A%
}^{\dagger }\hat{A}\hat{B}^{2}-2\frac{\tilde{g}_{0}}{\Delta _{-}}\hat{A}%
^{\dagger }\hat{A}^{2}\hat{B}e^{it\left( \delta _{\chi }-2\tilde{\delta}%
_{s}\right) }\right.  \nonumber \\
 &&+\frac{\tilde{g}_{0}^{2}}{\Delta _{-}^{2}}\hat{A}^{\dagger }\hat{A}%
^{3}e^{2it\left( \delta _{\chi }-2\tilde{\delta}_{s}\right) }+2\frac{\tilde{g%
}_{0}}{\Delta _{-}}\hat{A}^{\dagger }\hat{B}^{3}e^{-it\left( \delta _{\chi
}-2\tilde{\delta}_{s}\right) }+2\frac{\tilde{g}_{0}^{3}}{\Delta _{-}^{3}}%
\hat{A}^{3}\hat{B}^{\dagger }e^{3it\left( \delta _{\chi }-2\delta
_{s}\right) }  \nonumber \\
 &&\left. +4\frac{\tilde{g}_{0}}{\Delta _{-}}\hat{A}\hat{B}^{\dagger }%
\hat{B}^{2}e^{it\left( \delta _{\chi }-2\delta _{s}\right) }-5\frac{\tilde{g}%
_{0}^{2}}{\Delta _{-}^{2}}\hat{A}^{2}\hat{B}^{\dagger }\hat{B}e^{2it\left(
\delta _{\chi }-2\delta _{s}\right) }\right] +h.c.  \label{IDCE2}
\end{eqnarray}%
$\bullet $ for the mixed modulation frequency $\eta ^{(M)}\approx \Delta
_{+} $%
\begin{eqnarray}
\hat{H}_{NG}\left( \tilde{\varepsilon}_{g}\right) &=&i\frac{\tilde{g}_{0}%
}{4N}\frac{\tilde{\varepsilon}_{g}^{(M)}}{\tilde{g}_{0}}e^{-it\left[ \Delta
_{+}-2\tilde{\delta}_{+}-\delta _{\chi }-\eta ^{(M)}\right] }\left[ \hat{A}%
\hat{B}^{\dagger }\hat{B}^{2}-\frac{\tilde{g}_{0}}{\Delta _{-}}\hat{A}%
^{\dagger }\hat{A}\hat{B}^{2}e^{-it\left( \delta _{\chi }-2\tilde{\delta}%
_{s}\right) }\right.  \label{MDCE} \\
 &&-\frac{\tilde{g}_{0}^{2}}{\Delta _{-}^{2}}\hat{A}^{\dagger }\hat{B}%
^{3}e^{-2it\left( \delta _{\chi }-2\tilde{\delta}_{s}\right) }-3\frac{\tilde{%
g}_{0}^{3}}{\Delta _{-}^{3}}\hat{A}^{\dagger }\hat{A}^{3}e^{it\left( \delta
_{\chi }-2\tilde{\delta}_{s}\right) }+\frac{\tilde{g}_{0}}{\Delta _{-}}\hat{B%
}^{\dagger }\hat{B}^{3}e^{-it\left( \delta _{\chi }-2\tilde{\delta}%
_{s}\right) }  \nonumber \\
 &&\left. +4\frac{\tilde{g}_{0}^{2}}{\Delta _{-}^{2}}\hat{A}^{\dagger }%
\hat{A}^{2}\hat{B}-2\frac{\tilde{g}_{0}}{\Delta _{-}}\hat{A}^{2}\hat{B}%
^{\dagger }\hat{B}e^{it\left( \delta _{\chi }-2\tilde{\delta}_{s}\right) }+%
\frac{\tilde{g}_{0}^{2}}{\Delta _{-}^{2}}\hat{A}^{3}\hat{B}^{\dagger
}e^{2it\left( \delta _{\chi }-2\tilde{\delta}_{s}\right) }\right] +h.c.
\nonumber
\end{eqnarray}%
$\bullet $ for the Anti-DCE modulation frequency $\eta ^{(A)}\approx 2\omega
_{0}+\Delta _{-}$%
\begin{eqnarray}
\hat{H}_{NG}\left( \tilde{\varepsilon}_{g}\right) &=&i\frac{\tilde{\delta}%
_{-}}{4N}\frac{\tilde{g}_{0}}{\Delta _{-}}\frac{\tilde{\varepsilon}_{g}^{(A)}%
}{\tilde{g}_{0}}e^{-it\left[ 2\omega _{0}+\Delta _{-}+4\tilde{\delta}_{-}-2%
\tilde{\delta}_{+}-3\delta _{\chi }-\eta ^{(A)}\right] }\left[ \hat{A}^{3}%
\hat{B}^{\dagger }e^{-4it\tilde{\delta}_{s}}\right.  \label{ADCE} \\
 &&+\frac{\tilde{g}_{0}}{\Delta _{-}}\left( 3\hat{B}^{\dagger }\hat{B}-%
\hat{A}^{\dagger }\hat{A}\right) \hat{A}^{2}e^{-it\left( \delta _{\chi }+2%
\tilde{\delta}_{s}\right) }+3\frac{\tilde{g}_{0}^{2}}{\Delta _{-}^{2}}\left(
\hat{A}\hat{B}^{\dagger }\hat{B}^{2}-\hat{A}^{\dagger }\hat{A}^{2}\hat{B}%
\right) e^{-2it\delta _{\chi }}  \nonumber \\
 &&\left. +\frac{\tilde{g}_{0}^{3}}{\Delta _{-}^{3}}\left( \hat{B}%
^{\dagger }\hat{B}-3\hat{A}^{\dagger }\hat{A}\right) \hat{B}^{2}e^{-it\left(
3\delta _{\chi }-2\tilde{\delta}_{s}\right) }-\frac{\tilde{g}_{0}^{4}}{%
\Delta _{-}^{4}}\hat{A}^{\dagger }\hat{B}^{3}e^{-4it\left( \delta _{\chi }-%
\tilde{\delta}_{s}\right) }\right] +h.c.  \nonumber
\end{eqnarray}%
$\bullet $ for the Anti-IDCE modulation frequency $\eta ^{(AI)}\approx
2\Omega _{0}-\Delta _{-}$%
\begin{eqnarray}
\hat{H}_{NG}\left( \tilde{\varepsilon}_{g}\right) &=&i\frac{\tilde{\delta}%
_{-}}{4N}\frac{\tilde{g}_{0}}{\Delta _{-}}\frac{\tilde{\varepsilon}%
_{g}^{(AI)}}{\tilde{g}_{0}}e^{-it\left[ 2\Omega _{0}-\Delta _{-}-4\tilde{%
\delta}_{-}-2\tilde{\delta}_{+}+\delta _{\chi }-\eta ^{(AI)}\right] }\left[ -%
\hat{A}^{\dagger }\hat{B}^{3}e^{4it\tilde{\delta}_{s}}\right.  \label{AIDCE}
\\
 &&+\frac{\tilde{g}_{0}}{\Delta _{-}}\left( 3\hat{A}^{\dagger }\hat{A}-%
\hat{B}^{\dagger }\hat{B}\right) \hat{B}^{2}e^{it\left( \delta _{\chi }+2%
\tilde{\delta}_{s}\right) }+3\frac{\tilde{g}_{0}^{2}}{\Delta _{-}^{2}}\left(
\hat{A}\hat{B}^{\dagger }\hat{B}^{2}-\hat{A}^{\dagger }\hat{A}^{2}\hat{B}%
\right) e^{2it\delta _{\chi }}  \nonumber \\
 &&\left. +\frac{\tilde{g}_{0}^{3}}{\Delta _{-}^{3}}\left( \hat{A}%
^{\dagger }\hat{A}-3\hat{B}^{\dagger }\hat{B}\right) \hat{A}^{2}e^{it\left(
3\delta _{\chi }-2\tilde{\delta}_{s}\right) }+\frac{\tilde{g}_{0}^{4}}{%
\Delta _{-}^{4}}\hat{A}^{3}\hat{B}^{\dagger }e^{4it\left( \delta _{\chi }-%
\tilde{\delta}_{s}\right) }\right] +h.c.  \nonumber
\end{eqnarray}

\section{Analytical results for $N=1$}

\label{circuitqed}

For $N=1$ we work in the Schr\"{o}dinger picture and expand the wavefunction
corresponding to the Hamiltonian (\ref{H1}) as \cite{JPA}
\begin{equation}
|\psi (t)\rangle =e^{-it\lambda _{0}}A_{0}(t)|\varphi _{0}\rangle
+\sum_{n=1}^{\infty }\sum_{\mathcal{S}=\pm }e^{-it\lambda _{n,\mathcal{S}%
}}A_{n,\mathcal{S}}(t)|\varphi _{n,\mathcal{S}}\rangle ~,
\end{equation}%
where $\lambda _{n,\mathcal{S}}$ and $|\varphi _{n,\mathcal{S}}\rangle $ are
the $n$-excitations eigenvalues and eigenstates of the bare Jaynes-Cummings
Hamiltonian
\begin{equation}
\hat{H}_{JC}=\omega _{0}\hat{n}+\Omega _{0}|e\rangle \langle e|+g_{0}(\hat{a}%
\hat{\sigma}_{+}+\hat{a}^{\dagger }\hat{\sigma}_{-})~.
\end{equation}%
The index $\mathcal{S}$ labels the different eigenstates with the same
number of excitations $n$.

The well known eigenfrequencies are%
\begin{equation}
\lambda _{0}=0~,~\lambda _{n>0,\mathcal{S}}=\omega _{0}n+\frac{\mathcal{S}%
\beta _{n}-\Delta _{-}}{2}~,~\beta _{n}=\sqrt{\Delta _{-}^{2}+4g_{0}^{2}n}~,~%
\mathcal{S}=\pm ~,  \label{betan}
\end{equation}%
where $\Delta _{-}=\omega _{0}-\Omega _{0}$ is the bare detuning. The
Jaynes-Cummings eigenstates, also known as \emph{dressed states}, are%
\begin{equation}
|\varphi _{0}\rangle =|g,0\rangle ~,~|\varphi _{n>0,\mathcal{S}}\rangle ={\rm s}%
_{n,\mathcal{S}}|g,n\rangle +{\rm c}_{n,\mathcal{S}}|e,n-1\rangle ~,
\label{dress}
\end{equation}%
where we introduced the notation%
\begin{equation}
{\rm s}_{n,+}=\sin \theta _{n},~{\rm s}_{n,-}=\cos \theta _{n},~{\rm c}_{n,+}=\cos
\theta _{n},~{\rm c}_{n,-}=-\sin \theta _{n}~
\end{equation}%
with%
\begin{equation}
\theta _{n>0}=\arctan \frac{\Delta _{-}+\beta _{n}}{2g_{0}\sqrt{n}}~.
\end{equation}

We introduce new time-dependent probability amplitudes $b(t)$ via the
relations%
\begin{equation}
A_{0}(t)=e^{-it\left( \nu _{0}^{(1)}+\nu _{0}^{(2)}\right) }b_{0}(t)
\label{testa1}
\end{equation}%
\begin{eqnarray}
 A_{m,\mathcal{T}}(t) &=&\left\{ e^{-it\left( \nu _{m,\mathcal{T}%
}^{(1)}+\nu _{m,\mathcal{T}}^{(2)}\right) }b_{m,\mathcal{T}}(t)-\frac{1}{2i}%
\sum\nolimits_{j}^{\prime }\sum_{k=\omega ,\Omega ,g}\Pi _{m,\mathcal{T},-%
\mathcal{T}}^{k,j}e^{-it\left( \nu _{m,-\mathcal{T}}^{(1)}+\nu _{m,-\mathcal{%
T}}^{(2)}\right) }\right.  \nonumber \\
&&\left. \times \left[ \frac{e^{it\left( \lambda _{m,\mathcal{T}}-\lambda
_{m,-\mathcal{T}}+\eta ^{(j)}\right) }-1}{\left( \lambda _{m,\mathcal{T}%
}-\lambda _{m,-\mathcal{T}}+\eta ^{(j)}\right) }e^{i\phi _{k}^{(j)}}-\frac{%
e^{it\left( \lambda _{m,\mathcal{T}}-\lambda _{m,-\mathcal{T}}-\eta
^{(j)}\right) }-1}{\left( \lambda _{m,\mathcal{T}}-\lambda _{m,-\mathcal{T}%
}-\eta ^{(j)}\right) }e^{-i\phi _{k}^{(j)}}\right] b_{m,-\mathcal{T}%
}(t)\right\}  \nonumber \\
&&\times \exp \left[ i\sum\nolimits_{j}^{\prime }\sum_{k=\omega ,\Omega ,g}%
\frac{\Pi _{m,\mathcal{T},\mathcal{T}}^{k,j}}{\eta ^{(j)}}\left[ \cos \left(
\eta ^{(j)}t+\phi _{k}^{(j)}\right) -\cos \phi _{k}^{(j)}\right] \right] ~,
\label{testa2}
\end{eqnarray}%
where the sum $\sum\nolimits_{j}^{\prime }$ runs over \textquotedblleft
high\textquotedblright\ modulation frequencies $\eta ^{(j\prime )}\gtrsim
\omega _{0}$ and we defined the time-independent coefficients
\begin{equation}
\Pi _{m,\mathcal{T},\mathcal{S}}^{\omega ,j}\equiv \varepsilon _{\omega
}w_{\omega }^{(j)}\langle \varphi _{m,\mathcal{T}}|\hat{n}|\varphi _{m,%
\mathcal{S}}\rangle
\end{equation}%
\begin{equation}
\Pi _{m,\mathcal{T},\mathcal{S}}^{\Omega ,j}\equiv \varepsilon _{\Omega
}w_{\Omega }^{(j)}\langle \varphi _{m,\mathcal{T}}|e\rangle \langle
e|\varphi _{m,\mathcal{S}}\rangle
\end{equation}%
\begin{equation}
\Pi _{m,\mathcal{T},\mathcal{S}}^{g,j}\equiv \varepsilon
_{g}w_{g}^{(j)}\langle \varphi _{m,\mathcal{T}}|(\hat{a}\hat{\sigma}_{+}+%
\hat{a}^{\dagger }\hat{\sigma}_{-})|\varphi _{m,\mathcal{S}}\rangle ~.
\end{equation}%
These quantities are calculated in a straightforward manner using the
dressed states.

In equations (\ref{testa1}) -- (\ref{testa2}) we introduced small
\textquotedblleft intrinsic frequency shifts\textquotedblright\ \cite{JPA}
due to the elimination of the rapidly rotating terms throughout the
derivation:%
\begin{equation}
\nu _{0}^{(1)}=-\frac{1}{4}\sum_{j}\left\vert \varepsilon
_{d}^{(j)}\right\vert ^{2}\sum_{\mathcal{S}=\pm }\frac{\left\vert {\rm s}_{1,%
\mathcal{S}}\right\vert ^{2}}{\lambda _{1,\mathcal{S}}+\eta ^{(j)}}
\label{natal1}
\end{equation}%
\begin{equation}
\nu _{1,\mathcal{T}}^{(1)}=\frac{1}{4}\sum_{j}\left\vert \varepsilon
_{d}^{(j)}\right\vert ^{2}\sum_{\mathcal{S}=\pm }\left[ \frac{L_{1,1,%
\mathcal{S},\mathcal{T}}^{2}}{\lambda _{m,\mathcal{T}}+\eta ^{(j)}}-\frac{%
L_{1,m+1,\mathcal{T},\mathcal{S}}^{2}}{\lambda _{m+1,\mathcal{S}}-\lambda
_{m,\mathcal{T}}+\eta ^{(j)}}\right]
\end{equation}%
\begin{equation}
\nu _{m>1,\mathcal{T}}^{(1)}=\frac{1}{4}\sum_{j}\left\vert \varepsilon
_{d}^{(j)}\right\vert ^{2}\sum_{\mathcal{S}=\pm }\left[ \frac{L_{1,m,%
\mathcal{S},\mathcal{T}}^{2}}{\lambda _{m,\mathcal{T}}-\lambda _{m-1,%
\mathcal{S}}+\eta ^{(j)}}-\frac{L_{1,m+1,\mathcal{T},\mathcal{S}}^{2}}{%
\lambda _{m+1,\mathcal{S}}-\lambda _{m,\mathcal{T}}+\eta ^{(j)}}\right]
\end{equation}%
\begin{equation}
\nu _{0}^{(2)}=-\sum_{\mathcal{S}=\pm }\frac{{\rm c}_{2,\mathcal{S}%
}^{2}g_{0}^{2}+2{\rm s}_{2,\mathcal{S}}^{2}\chi _{0}^{2}}{\lambda _{2,\mathcal{S%
}}}~,~\nu _{1,\mathcal{T}}^{(2)}=-\sum_{\mathcal{S}=\pm }\frac{\Lambda _{m+2,%
\mathcal{T},\mathcal{S}}^{2}g_{0}^{2}+L_{2,m+2,\mathcal{T},\mathcal{S}%
}^{2}\chi _{0}^{2}}{\lambda _{m+2,\mathcal{S}}-\lambda _{m,\mathcal{T}}}
\end{equation}%
\begin{equation}
\nu _{2,\mathcal{T}}^{(2)}=\sum_{\mathcal{S}=\pm }\left[ \frac{\Lambda _{2,%
\mathcal{S},\mathcal{T}}^{2}g_{0}^{2}+L_{2,2,\mathcal{S},\mathcal{T}%
}^{2}\chi _{0}^{2}}{\lambda _{m,\mathcal{T}}}-\frac{\Lambda _{m+2,\mathcal{T}%
,\mathcal{S}}^{2}g_{0}^{2}+L_{2,m+2,\mathcal{T},\mathcal{S}}^{2}\chi _{0}^{2}%
}{\lambda _{m+2,\mathcal{S}}-\lambda _{m,\mathcal{T}}}\right]
\end{equation}%
\begin{equation}
\nu _{m>2,\mathcal{T}}^{(2)}=\sum_{\mathcal{S}=\pm }\left[ \frac{\Lambda _{m,%
\mathcal{S},\mathcal{T}}^{2}g_{0}^{2}+L_{2,m,\mathcal{S},\mathcal{T}%
}^{2}\chi _{0}^{2}}{\lambda _{m,\mathcal{T}}-\lambda _{m-2,\mathcal{S}}}-%
\frac{\Lambda _{m+2,\mathcal{T},\mathcal{S}}^{2}g_{0}^{2}+L_{2,m+2,\mathcal{T%
},\mathcal{S}}^{2}\chi _{0}^{2}}{\lambda _{m+2,\mathcal{S}}-\lambda _{m,%
\mathcal{T}}}\right] ~,  \label{natal2}
\end{equation}%
where we defined%
\begin{equation}
\Lambda _{m+2,\mathcal{T},\mathcal{S}}\equiv \langle \varphi _{m,\mathcal{T}%
}|\hat{a}\hat{\sigma}_{-}|\varphi _{m+2,\mathcal{S}}\rangle ~,~\quad
L_{k,m+k,\mathcal{T},\mathcal{S}}\equiv \langle \varphi _{m,\mathcal{T}}|%
\hat{a}^{k}|\varphi _{m+k,\mathcal{S}}\rangle ~.
\end{equation}

The new probability amplitudes obey the differential equations (to simplify
the notation we denote $b_{0,\mathcal{T}}\equiv b_{0}$, $|\varphi _{0,%
\mathcal{T}}\rangle \equiv |\varphi _{0}\rangle $ and $\lambda _{0,\mathcal{T%
}}\equiv \lambda _{0}$)%
\begin{eqnarray}
\dot{b}_{m,\mathcal{T}} &=&-i\sum_{\mathcal{S}}\sum\nolimits_{j}^{\prime
\prime }\sum_{k=\omega ,\Omega ,g}\Pi _{m,\mathcal{T},\mathcal{S}%
}^{k,j}e^{it\left( \bar{\lambda}_{m,\mathcal{T}}-\bar{\lambda}_{m,\mathcal{S}%
}\right) }\sin (\eta ^{(j)}t+\phi _{k}^{(j)})b_{m,\mathcal{S}}  \nonumber \\
&&+\sum_{\mathcal{S}}\sum\nolimits_{j}^{\prime }\left[ \Theta _{m+2,\mathcal{%
T},\mathcal{S}}^{(j)}e^{-it\left( \bar{\lambda}_{m+2,\mathcal{S}}-\bar{%
\lambda}_{m,\mathcal{T}}-\eta ^{(j)}\right) }b_{m+2,\mathcal{S}}\right.
\nonumber \\
&&\left. -\Theta _{m,\mathcal{S},\mathcal{T}}^{(j)\ast }e^{it\left( \bar{%
\lambda}_{m,\mathcal{T}}-\bar{\lambda}_{m-2,\mathcal{S}}-\eta ^{(j)}\right)
}b_{m-2,\mathcal{S}}\right]  \nonumber \\
&&+\frac{i}{2}\sum_{\mathcal{S}}\sum_{j}\left[ \varepsilon
_{d}^{(j)}e^{-it\left( \bar{\lambda}_{m+1,\mathcal{S}}-\bar{\lambda}_{m,%
\mathcal{T}}-\eta ^{(j)}\right) }L_{1,m+1,\mathcal{T},\mathcal{S}}b_{m+1,%
\mathcal{S}}\right.  \nonumber \\
&&\left. +\varepsilon _{d}^{(j)\ast }e^{it\left( \bar{\lambda}_{m,\mathcal{T}%
}-\bar{\lambda}_{m-1,\mathcal{S}}-\eta ^{(j)}\right) }L_{1,m,\mathcal{S},%
\mathcal{T}}^{\ast }b_{m-1,\mathcal{S}}\right]  \label{b}
\end{eqnarray}%
with time-independent coefficients (where $m>0$)%
\begin{equation}
\Theta _{2,\mathcal{T},\mathcal{S}}^{(j)}=\frac{1}{2}\sum_{\mathcal{R=\pm }%
}\sum_{l=\omega ,\Omega ,g}\frac{g_{0}\Lambda _{2,\mathcal{T},\mathcal{R}%
}-i\chi _{0}L_{2,2,\mathcal{T},\mathcal{R}}}{\lambda _{2,\mathcal{R}%
}-\lambda _{2,\mathcal{S}}+\eta ^{(j)}}\Pi _{2,\mathcal{R},\mathcal{S}%
}^{l,j}e^{i\phi _{l}^{(j)}}-\frac{1}{2}\left( \varepsilon _{g}^{(j)}\Lambda
_{2,\mathcal{T},\mathcal{S}}-i\varepsilon _{\chi }^{(j)}L_{2,2,\mathcal{T},%
\mathcal{S}}\right)  \label{teta1}
\end{equation}%
\begin{eqnarray}
\Theta _{m+2,\mathcal{T},\mathcal{S}}^{(j)} &=&\frac{1}{2}\sum_{\mathcal{R%
}=\pm }\sum_{l=\omega ,\Omega ,g}\left[ \frac{g_{0}\Lambda _{m+2,\mathcal{T},%
\mathcal{R}}-i\chi _{0}L_{2,m+2,\mathcal{T},\mathcal{R}}}{\lambda _{m+2,%
\mathcal{R}}-\lambda _{m+2,\mathcal{S}}+\eta ^{(j)}}\Pi _{m+2,\mathcal{R},%
\mathcal{S}}^{l,j}e^{i\phi _{l}^{(j)}}\right.  \label{teta2} \\
&& \quad \left. -\frac{g_{0}\Lambda _{m+2,\mathcal{R},\mathcal{S}}-i\chi
_{0}L_{2,m+2,\mathcal{R},\mathcal{S}}}{\lambda _{m,\mathcal{T}}-\lambda _{m,%
\mathcal{R}}+\eta ^{(j)}}\Pi _{m,\mathcal{T},\mathcal{R}}^{l,j}e^{i\phi
_{l}^{(j)}}\right] -\frac{1}{2}\left( \varepsilon _{g}^{(j)}\Lambda _{m+2,%
\mathcal{T},\mathcal{S}}-i\varepsilon _{\chi }^{(j)}L_{2,m+2,\mathcal{T},%
\mathcal{S}}\right) ~.  \nonumber
\end{eqnarray}%
Equation (\ref{b}) was deduced under the following approximations [recall
that $(j^{\prime })$ stands for \textquotedblleft high\textquotedblright\
modulation frequencies $\eta ^{(j\prime )}\gtrsim \omega _{0}$ and $k=\omega
,\Omega ,g$]%
\begin{equation}
\frac{\left\vert \Pi _{m,\mathcal{S},\mathcal{S}}^{k,j\prime }-\Pi _{m,%
\mathcal{-S},\mathcal{-S}}^{k,j\prime }\right\vert }{\omega _{0}},\frac{%
\left\vert \Pi _{m,\mathcal{S},-\mathcal{S}}^{k,j\prime }\right\vert }{%
\omega _{0}},\frac{\left\vert \Pi _{m\pm 2,\mathcal{S},\mathcal{S}%
}^{k,j\prime }-\Pi _{m,\mathcal{-S},\mathcal{-S}}^{k,j\prime }\right\vert }{%
\omega _{0}},\frac{\left\vert \Pi _{m\pm 1,\mathcal{S},\mathcal{S}%
}^{k,j\prime }-\Pi _{m,\mathcal{-S},\mathcal{-S}}^{k,j\prime }\right\vert }{%
\omega _{0}}\ll 1  \label{ap1}
\end{equation}%
\begin{equation}
\frac{\{g_{0},\varepsilon _{g}\}\Lambda _{m+2,\mathcal{T},\mathcal{S}}}{%
\omega _{0}},\frac{\left\{ \chi _{0},\varepsilon _{\chi }\right\} L_{2,m,%
\mathcal{T},\mathcal{S}}}{\omega _{0}},\frac{\varepsilon _{d}L_{1,m+1,%
\mathcal{T},\mathcal{S}}}{\omega _{0}}\ll 1~.  \label{ap2}
\end{equation}

Notice that in equation (\ref{b}) the resonant modulation frequencies $\eta
^{(j)}$ correspond to the difference between two \textquotedblleft
corrected\textquotedblright\ eigenfrequencies defined as%
\begin{equation}
\bar{\lambda}_{m,\mathcal{T}}\equiv \lambda _{m,\mathcal{T}}+\nu _{m,%
\mathcal{T}}^{(1)}+\nu _{m,\mathcal{T}}^{(2)}~  \label{lamb}
\end{equation}%
(we denote $\bar{\lambda}_{0,\mathcal{T}}\equiv \bar{\lambda}_{0}$). So the
Jaynes-Cummings eigenfrequencies are corrected by the frequency shifts $\nu
_{m,\mathcal{T}}^{(1)}$ and $\nu _{m,\mathcal{T}}^{(2)}$. In equation (\ref%
{lamb}) we neglected the additional frequency shift $\nu _{m,\mathcal{T}%
}^{(3)}$ due to the modulation depths $\varepsilon _{g}$, $\varepsilon
_{\omega }$, $\varepsilon _{\Omega }$ and $\varepsilon _{\chi }$, of the
order%
\begin{equation}
O(\nu _{m,\mathcal{T}}^{(3)})\sim \frac{m\varepsilon _{\chi }^{2}}{\omega
_{0}},\frac{\left( \Pi _{m,\mathcal{S},\mathcal{-S}}^{k,j\prime }\right) ^{2}%
}{\omega _{0}},\frac{\left( \Pi _{m\pm 2,\mathcal{S},\mathcal{S}}^{k,j\prime
}-\Pi _{m,\mathcal{-S},\mathcal{-S}}^{k,j\prime }\right) ^{2}}{\omega _{0}}%
\mbox{ for }k=\omega ,\Omega ,g~.  \label{ap3}
\end{equation}%
We call these neglected frequency shifts \textquotedblleft Systematic-error
frequency shifts\textquotedblright\ (SEFS), since they appear due to the
systematic simplification of the differential equations for $b_{m,\mathcal{T}%
}$ using the RWA \cite{JPA}. The knowledge of SEFS is important because they
slightly alter the resonant modulation frequencies, so ultimately they must
be found numerically or experimentally.

\subsection{Simplified formulae in the resonant regime}

\label{lemmy}

For $\Delta _{-}=0$ we obtain the expressions%
\begin{equation}
\bar{\lambda}_{0}=-\frac{1}{4}\sum_{j}\frac{|\varepsilon _{d}^{(j)}|^{2}}{%
\omega _{0}+\eta ^{(j)}}-\left( \delta _{+}+\frac{1}{2}\delta _{\chi
}\right) ~~
\end{equation}%
\begin{equation}
\bar{\lambda}_{m>0,\mathcal{S}}=\omega _{0}m+\mathcal{S}g_{0}\sqrt{m}-\left(
\delta _{+}+m\delta _{\chi }\right) -\frac{1}{4}\sum_{j}\frac{|\varepsilon
_{d}^{(j)}|^{2}}{\omega _{0}+\eta ^{(j)}}~,~
\end{equation}%
\begin{equation}
\delta _{\pm }=\frac{g_{0}^{2}}{\Delta _{\pm }}~,~\delta _{\chi }=\frac{%
4\chi _{0}^{2}}{\Delta _{+}}~
\end{equation}%
\begin{equation}
|\varphi _{m,\mathcal{S}}\rangle =\frac{1}{\sqrt{2}}\left( |g,m\rangle +%
\mathcal{S}|e,m-1\rangle \right) ~.
\end{equation}%
The coefficients are (for $m>0$)

\begin{equation}
\Theta _{2,\mathcal{T},\mathcal{+}}^{(j)}=\frac{1}{4}\sqrt{2}\left[ \left(
g_{0}-2i\chi _{0}\sqrt{2}\right) \frac{\varepsilon _{\omega }^{(j)}}{\eta
^{(j)}}+g_{0}\frac{\varepsilon _{\Omega }^{(j)}}{\eta ^{(j)}}-\left( 1+\frac{%
2i\chi _{0}}{\eta ^{(j)}}\right) \varepsilon _{g}^{(j)}+i\sqrt{2}\varepsilon
_{\chi }^{(j)}\right]
\end{equation}%
\begin{equation}
\Theta _{2,\mathcal{T},\mathcal{-}}^{(j)}=-\frac{1}{4}\sqrt{2}\left[ \left(
g_{0}+2i\chi _{0}\sqrt{2}\right) \frac{\varepsilon _{\omega }^{(j)}}{\eta
^{(j)}}+g_{0}\frac{\varepsilon _{\Omega }^{(j)}}{\eta ^{(j)}}-\left( 1+\frac{%
2i\chi _{0}}{\eta ^{(j)}}\right) \varepsilon _{g}^{(j)}-i\sqrt{2}\varepsilon
_{\chi }^{(j)}\right]
\end{equation}%
\begin{equation}
\Theta _{m+2,\mathcal{+},\mathcal{+}}^{(j)}=\frac{\sqrt{m+1}}{4}\left[
[g_{0}-2i\chi _{0}(\sqrt{m+2}+\sqrt{m})]\frac{\varepsilon _{\omega }^{(j)}}{%
\eta ^{(j)}}+g_{0}\frac{\varepsilon _{\Omega }^{(j)}}{\eta ^{(j)}}%
-\varepsilon _{g}^{(j)}+i\varepsilon _{\chi }^{(j)}(\sqrt{m+2}+\sqrt{m})%
\right]
\end{equation}%
\begin{equation}
\Theta _{m+2,\mathcal{-},\mathcal{-}}^{(j)}=\frac{\sqrt{m+1}}{4}\left[
-[g_{0}+2i\chi _{0}(\sqrt{m+2}+\sqrt{m})]\frac{\varepsilon _{\omega }^{(j)}}{%
\eta ^{(j)}}-g_{0}\frac{\varepsilon _{\Omega }^{(j)}}{\eta ^{(j)}}%
+\varepsilon _{g}^{(j)}+i\varepsilon _{\chi }^{(j)}(\sqrt{m+2}+\sqrt{m})%
\right]
\end{equation}%
\begin{equation}
\Theta _{m+2,\mathcal{+},\mathcal{-}}^{(j)}=\frac{\sqrt{m+1}}{4}\left[
-[g_{0}+2i\chi _{0}(\sqrt{m+2}-\sqrt{m})]\frac{\varepsilon _{\omega }^{(j)}}{%
\eta ^{(j)}}-g_{0}\frac{\varepsilon _{\Omega }^{(j)}}{\eta ^{(j)}}%
+\varepsilon _{g}^{(j)}+i\varepsilon _{\chi }^{(j)}(\sqrt{m+2}-\sqrt{m})%
\right]
\end{equation}%
\begin{equation}
\Theta _{m+2,\mathcal{-},\mathcal{+}}^{(j)}=\frac{\sqrt{m+1}}{4}\left[
[g_{0}-2i\chi _{0}(\sqrt{m+2}-\sqrt{m})]\frac{\varepsilon _{\omega }^{(j)}}{%
\eta ^{(j)}}+g_{0}\frac{\varepsilon _{\Omega }^{(j)}}{\eta ^{(j)}}%
-\varepsilon _{g}^{(j)}+i\varepsilon _{\chi }^{(j)}(\sqrt{m+2}-\sqrt{m})%
\right] ~
\end{equation}%
\begin{equation}
L_{1,1,\mathcal{T},\mathcal{S}}=\frac{1}{\sqrt{2}}~,~L_{1,m+1,\mathcal{T},%
\mathcal{T}}=\frac{1}{2}\left( \sqrt{m+1}+\sqrt{m}\right) ~,~L_{1,m+1,%
\mathcal{T},\mathcal{-T}}=\frac{1}{2}\left( \sqrt{m+1}-\sqrt{m}\right) ~.
\end{equation}%
From equations (\ref{ap1}), (\ref{ap2}) and (\ref{ap3}) we derive explicitly
the underlying approximations and SEFS in the resonant regime%
\begin{equation}
\varepsilon _{\omega },\varepsilon _{\Omega },g_{0}\sqrt{M},\varepsilon _{g}%
\sqrt{M},\varepsilon _{d}\sqrt{M},\chi _{0}M,\varepsilon _{\chi }M\ll \omega
_{0}
\end{equation}%
\begin{equation}
O(\nu _{m,\mathcal{T}}^{(3)})\sim \frac{\varepsilon _{\omega }^{2}}{\omega
_{0}},\frac{\varepsilon _{\Omega }^{2}}{\omega _{0}},\frac{m\varepsilon
_{g}^{2}}{\omega _{0}},\frac{m\varepsilon _{\chi }^{2}}{\omega _{0}}~.
\end{equation}

\subsection{Simplified formulae in the dispersive regime}

\label{cirdisp}

For $|\Delta _{-}|/2\gg g_{0}\sqrt{n}$ we obtain after expanding $\beta _{n}$
in equation (\ref{betan})%
\begin{equation}
\lambda _{n,\mathcal{D}}\simeq \omega _{0}n+\allowbreak \delta _{-}n-\alpha
n^{2}~,~\lambda _{n,\mathcal{-D}}\simeq \omega _{0}n-\Delta _{-}-\delta
_{-}n+\alpha n^{2}
\end{equation}%
\begin{equation}
|\varphi _{m,\mathcal{D}}\rangle \simeq \left( |\mathbf{g},m\rangle +\frac{%
g_{0}}{\Delta _{-}}\sqrt{m}|\mathbf{e},m-1\rangle \right) ~,~|\varphi _{m,%
\mathcal{-D}}\rangle \simeq -\mathcal{D}\left( |\mathbf{e},m-1\rangle -\frac{%
g_{0}}{\Delta _{-}}\sqrt{m}|\mathbf{g},m\rangle \right) ,
\end{equation}%
where $\mathcal{D}$ is the \textquotedblleft detuning
symbol\textquotedblright , equation (\ref{dsym}), and the effective Kerr
nonlinearity strength is $\alpha =g_{0}^{4}/\Delta _{-}^{3}~.$The
coefficients are (for $m>0$)%
\begin{equation}
\Theta _{2,\mathcal{T},\mathcal{D}}^{(j)}=\frac{1}{2}\sqrt{2}\delta _{-}%
\left[ \left( \frac{\eta ^{(j)}-2\Delta _{-}}{\eta ^{(j)}-\Delta _{-}}-\frac{%
2i\chi _{0}}{\delta _{-}}\right) \frac{\varepsilon _{\omega }^{(j)}}{\eta
^{(j)}}+\frac{\varepsilon _{\Omega }^{(j)}}{\eta ^{(j)}-\Delta _{-}}+\frac{%
(2\Delta _{-}-\eta ^{(j)})}{(\eta ^{(j)}-\Delta _{-})}\frac{\varepsilon
_{g}^{(j)}}{g_{0}}+\frac{i\varepsilon _{\chi }^{(j)}}{\delta _{-}}\right]
\end{equation}%
\begin{equation}
\Theta _{2,\mathcal{T},\mathcal{-D}}^{(j)}=\frac{1}{2}g_{0}\mathcal{D}\left[
-\left( 1+i\frac{2\chi _{0}}{\Delta _{-}}\frac{2\eta ^{(j)}+\Delta _{-}}{%
\eta ^{(j)}+\Delta _{-}}\right) \frac{\varepsilon _{\omega }^{(j)}}{\eta
^{(j)}}-\frac{\varepsilon _{\Omega }^{(j)}}{\eta ^{(j)}}+\frac{\varepsilon
_{g}^{(j)}}{g_{0}}+\frac{2i\varepsilon _{\chi }^{(j)}}{\Delta _{-}}\right]
\end{equation}%
\begin{equation}
\Theta _{m+2,\mathcal{D},\mathcal{-D}}^{(j)}=\frac{1}{2}g_{0}\mathcal{D}%
\sqrt{m+1}\left[ -\left( 1+i\frac{2\chi _{0}}{\Delta _{-}}\frac{2\eta
^{(j)}+\Delta _{-}}{\eta ^{(j)}+\Delta _{-}}\right) \frac{\varepsilon
_{\omega }^{(j)}}{\eta ^{(j)}}-\frac{\varepsilon _{\Omega }^{(j)}}{\eta
^{(j)}}+\frac{\varepsilon _{g}^{(j)}}{g_{0}}+\frac{2i\varepsilon _{\chi
}^{(j)}}{\Delta _{-}}\right]
\end{equation}%
\begin{equation}
\Theta _{m+2,\mathcal{-D},\mathcal{D}}^{(j)}=\frac{\mathcal{D}\delta
_{-}g_{0}\sqrt{m(m+1)(m+2)}}{2\Delta _{-}}\left[ \frac{\eta ^{(j)}-3\Delta
_{-}}{\eta ^{(j)}-\Delta _{-}}\frac{\varepsilon _{\omega }^{(j)}}{\eta ^{(j)}%
}+\frac{\eta ^{(j)}+\Delta _{-}}{\eta ^{(j)}-\Delta _{-}}\frac{\varepsilon
_{\Omega }^{(j)}}{\eta ^{(j)}}-\frac{\eta ^{(j)}-3\Delta _{-}}{\eta
^{(j)}-\Delta _{-}}\frac{\varepsilon _{g}^{(j)}}{g_{0}}\right]
\end{equation}%
\begin{equation}
\Theta _{m+2,\mathcal{D},\mathcal{D}}^{(j)}=\frac{\delta _{-}\sqrt{(m+1)(m+2)%
}}{2}\left[ \left( \frac{\eta ^{(j)}-2\Delta _{-}}{\eta ^{(j)}-\Delta _{-}}-%
\frac{2i\chi _{0}}{\delta _{-}}\right) \frac{\varepsilon _{\omega }^{(j)}}{%
\eta ^{(j)}}+\frac{\varepsilon _{\Omega }^{(j)}}{\eta ^{(j)}-\Delta _{-}}-%
\frac{\eta ^{(j)}-2\Delta _{-}}{\eta ^{(j)}-\Delta _{-}}\frac{\varepsilon
_{g}^{(j)}}{g_{0}}+\frac{i\varepsilon _{\chi }^{(j)}}{\delta _{-}}\right]
\end{equation}%
\begin{equation}
\Theta _{m+2,\mathcal{-D},\mathcal{-D}}^{(j)}=\frac{\delta _{-}\sqrt{m(m+1)}%
}{2}\left[ -\left( \frac{\eta ^{(j)}-2\Delta _{-}}{\eta ^{(j)}-\Delta _{-}}+%
\frac{2i\chi _{0}}{\delta _{-}}\right) \frac{\varepsilon _{\omega }^{(j)}}{%
\eta ^{(j)}}-\frac{\varepsilon _{\Omega }^{(j)}}{\eta ^{(j)}-\Delta _{-}}+%
\frac{\eta ^{(j)}-2\Delta _{-}}{\eta ^{(j)}-\Delta _{-}}\frac{\varepsilon
_{g}^{(j)}}{g_{0}}+\frac{i\varepsilon _{\chi }^{(j)}}{\delta _{-}}\right]
\end{equation}%
\begin{equation}
L_{1,m+1,\mathcal{D},\mathcal{D}}=\sqrt{m+1}~,~L_{1,m+1,\mathcal{D},\mathcal{%
-D}}=\frac{g_{0}}{\left\vert \Delta _{-}\right\vert }~,~L_{1,m+1,\mathcal{-D}%
,\mathcal{-D}}=\sqrt{m}~,~L_{1,m+1,\mathcal{-D},\mathcal{D}}\sim O\lbrack
(g_{0}/\Delta _{-})^{6}]~.
\end{equation}

The frequency shifts are%
\begin{equation}
\nu _{m\geq 0,\mathcal{T}}^{(1)}=-\frac{1}{4}\sum_{j}\frac{\left\vert
\varepsilon _{d}^{(j)}\right\vert ^{2}}{\omega _{0}+\eta ^{(j)}}~,~\quad \nu
_{0,\mathcal{T}}^{(2)}=-\left[ \delta _{+}+\frac{1}{2}\delta _{\chi }\right]
\end{equation}%
\begin{equation}
\nu _{m>0,\mathcal{D}}^{(2)}=-(m+1)\delta _{+}-\left( m+\frac{1}{2}\right)
\delta _{\chi }~,~\nu _{m>0,\mathcal{-D}}^{(2)}=(m-1)\delta _{+}-\left( m-%
\frac{1}{2}\right) \delta _{\chi }.
\end{equation}

From equations. (\ref{ap1}), (\ref{ap2}) and (\ref{ap3}) we derive
explicitly the underlying approximations and SEFS in the dispersive regime%
\begin{equation}
\varepsilon _{\omega },\varepsilon _{\Omega },g_{0}\sqrt{m},\varepsilon _{g}%
\sqrt{m},\varepsilon _{d}\sqrt{m},\chi _{0}m,\varepsilon _{\chi }m\ll \omega
_{0}
\end{equation}%
\begin{equation}
O(\nu _{m,\mathcal{T}}^{(3)})\sim \left( \frac{g_{0}\sqrt{m}}{\Delta _{-}}%
\right) ^{2}\frac{\varepsilon _{\omega }^{2}}{\omega _{0}},\left( \frac{g_{0}%
\sqrt{m}}{\Delta _{-}}\right) ^{2}\frac{\varepsilon _{\Omega }^{2}}{\omega
_{0}},\frac{m\varepsilon _{g}^{2}}{\omega _{0}},\frac{m\varepsilon _{\chi
}^{2}}{\omega _{0}}~.
\end{equation}

\begin{acknowledgments}
IMS acknowledges financial support by CAPES (Brazilian agency). AVD
acknowledges partial support by CNPq, Conselho Nacional de Desenvolvimento
Cient\'{\i}fico e Tecnol\'{o}gico -- Brazil.
\end{acknowledgments}


\begin{thebibliography}{99}
\bibitem{book} Dodonov V V 2001 \emph{Adv. Chem. Phys.} \textbf{119} 309

\bibitem{vdodonov} Dodonov V V 2010 \emph{Phys. Scr.} \textbf{82} 038105

\bibitem{revDal} Dalvit D A R, Maia Neto P A and Mazzitelli F D 2011 Casimir
Physics (Lecture Notes in Physics vol 834) ed D Dalvit, P Milonni, D Roberts
and F da Rosa (Berlin: Springer) p 419

\bibitem{nori} Nation P D \emph{et al.} 2012 \emph{Rev. Mod. Phys.} \textbf{%
84} 1

\bibitem{CAMOP-me} Dodonov A V 2013 \emph{Phys. Scr.} \textbf{87} 038103

\bibitem{2} Dodonov A V and Dodonov V V 2011 \emph{Phys. Lett.} A \textbf{375%
} 4261

\bibitem{bound2} Law C K 1994 \emph{Phys. Rev.} A \textbf{49} 433

\bibitem{diss4} Dodonov V V and Dodonov A V 2005 \emph{J. Russ. Laser Res.}
\textbf{26} 445

\bibitem{cir1} Blais A \emph{et al.} 2004 \emph{Phys. Rev.} A \textbf{69}
062320

\bibitem{cir2} Wallraff A \emph{et al.} 2004 \emph{Nature} \textbf{431} 162

\bibitem{cir3} Schoelkopf R J and Girvin S M 2008 \emph{Nature} \textbf{451}
664

\bibitem{meta} L\"{a}hteenm\"{a}ki P \emph{et al.} 2013 \emph{Proc. Nat.
Acad. Sci.} \textbf{110} 4234

\bibitem{sat1} Lambrecht A, Jaekel M -T and Reynaud S 1996 \emph{Phys. Rev.
Lett.} \textbf{77} 615

\bibitem{sat2} Dezael F X and Lambrecht A 2010 \emph{Europhys. Lett.}
\textbf{89} 14001

\bibitem{diss1} Dodonov V V 1998 \emph{Phys. Rev.} A \textbf{58} 4147

\bibitem{diss2} Schaller G \emph{et al.} 2002 \emph{Phys. Lett.} A \textbf{%
297} 81

\bibitem{diss3} Schaller G \emph{et al.} 2002 \emph{Phys. Rev.} A \textbf{66}
023812

\bibitem{diss5} Dodonov V V and Dodonov A V 2006 \emph{J. Phys.} B \textbf{39%
} S749

\bibitem{diss6} Dodonov V V 2009 \emph{Phys. Rev.} A \textbf{80} 023814

\bibitem{asym1} Saito H and Hyuga H 2002 \emph{Phys. Rev.} A \textbf{65}
053804

\bibitem{asym2} Lombardo F C and Mazzitelli F D 2010 \emph{Phys. Scr.}
\textbf{82} 038113

\bibitem{bound1} Moore G T 1970 \emph{J. Math. Phys.} \textbf{11} 2679

\bibitem{bound3} Mundarain D F and Maia Neto P A 1998 \emph{Phys. Rev.} A
\textbf{57} 1379

\bibitem{bound4} Montarezi M and Miri M 2005 \emph{Phys. Rev.} A \textbf{71}
063814

\bibitem{bound5} Johansson J R \emph{et al.} 2009 \emph{Phys. Rev. Lett.}
\textbf{103} 147003

\bibitem{bound6} Fosco C D, Lombardo F C and Mazzitelli F D 2013 \emph{Phys.
Rev.} D \textbf{87} 105008

\bibitem{bound7} Rego A L C \emph{et al.} 2014 \emph{Phys. Rev.} D \textbf{90%
} 025003

\bibitem{JPA} Dodonov A V 2013 \emph{J. Phys.} A \textbf{47} 285303

\bibitem{schleich} Schleich W P 2001 \emph{Quantum Optics in Phase Space}
(Berlin: Wiley)

\bibitem{zeilinger} Fujii T \emph{et al.} 2011 \emph{Phys. Rev.} B \textbf{84%
} 174521

\bibitem{seed} Faccio D and Carusotto I 2011 \emph{Europhys. Lett.} \textbf{%
96} 24006

\bibitem{polar1} Andr\'{e} A \emph{et al.} 2006 \emph{Nat. Phys.} \textbf{2}
636

\bibitem{polar2} Carr L D \emph{et al.} 2009 \emph{New J. Phys.} \textbf{11}%
, 055049

\bibitem{cir4} Clarke J and Wilhelm F K 2008 \emph{Nature} \textbf{452}, 1031

\bibitem{cir5} Fink J M \emph{et al.} 2009 \emph{Phys. Rev. Lett.} \textbf{%
103} 083601

\bibitem{vogel} Vogel W and Welsch D -G 2006 \emph{Quantum Optics} (Berlin:
Wiley)

\bibitem{HP} Garraway B M 2011 \emph{Phil. Trans. R. Soc.} A \textbf{369}
1137

\bibitem{kerr1} Gerry C C and Rodrigues S 1987 \emph{Phys. Rev.} A \textbf{36%
} 5444

\bibitem{milburn1} Milburn G J 1990 \emph{Phys. Rev.} A \textbf{41} 6567

\bibitem{milburn2} Milburn G J and Holmes C A 1991 \emph{Phys. Rev.} A
\textbf{44} 4704

\bibitem{kerr2} Gerry C C, Grobe R and Vrscay E R 1991 \emph{Phys. Rev.} A
\textbf{43} 361

\bibitem{milburn3} Wielinga B and Milburn G J 1992 \emph{Phys. Rev.} A
\textbf{46} 762

\bibitem{milburn4} Wielinga B and Milburn G J 1993 \emph{Phys. Rev.} A
\textbf{48} 2494

\bibitem{milburn5} Wielinga B and Milburn G J 1994 \emph{Phys. Rev.} A
\textbf{49} 5042

\bibitem{kryuch2} Kryuchkyan G Yu \emph{et al.} 1995 \emph{Quantum
Semiclass. Opt.} \textbf{7} 965

\bibitem{kryuch1} Kryuchkyan G Yu and Kheruntsyan K V 1996 \emph{Opt. Commun.%
} \textbf{127} 230

\bibitem{leonski} Leo\'nski W 1996 \emph{Phys. Rev.} A \textbf{54} 3369

\bibitem{kerr3} Lisowski T 1997 \emph{Quantum Semiclass. Opt.} \textbf{9} 103

\bibitem{kryuch3} Gevorgyan T V and Kryuchkyan G Yu 2013 \emph{J. Mod. Opt.}
\textbf{60} 860

\bibitem{back1} Nagatani Y and Shigetomi K 2000 \emph{Phys. Rev.} A \textbf{%
62} 022117

\bibitem{back2} Carusotto I \emph{et al.} 2012 \emph{Phys. Rev.} A \textbf{85%
} 023805

\bibitem{puri} Puri R R 2001 \emph{Mathematical Methods of Quantum Optics}
(Springer, Berlin).

\bibitem{1} Dodonov A V \emph{et al.} 2011 \emph{J. Phys.} B \textbf{44}
225502

\bibitem{bea} Beaudoin F, Gambetta J M and Blais A 2011 \emph{Phys. Rev.} A
\textbf{84} 043832

\bibitem{pra} Dodonov V V and Klimov A B 1996 \emph{Phys. Rev.} A \textbf{53}
2664

\bibitem{hyper} Mizrahi S S and Dodonov V V 2002 \emph{J. Phys.} A \textbf{35%
} 8847

\bibitem{sinaia1} Dodonov A V and Dodonov V V 2012 \emph{Phys. Rev.} A
\textbf{86}, 015801

\bibitem{sinaia2} Dodonov A V and Dodonov V V 2013 \emph{Phys. Scr.} T
\textbf{153} 014017

\bibitem{pla} Dodonov V V 1995 \emph{Phys. Lett.} A \textbf{207} 126

\bibitem{pla1} de Castro A S M and Dodonov V V 2013 \emph{J. Phys.} A
\textbf{46} 395304

\bibitem{pla2} de Castro A S M, Cacheffo A and Dodonov V V 2013 \emph{Phys.
Rev.} A \textbf{87} 033809

\bibitem{pla3} de Castro A S M, Cacheffo A and Dodonov V V 2014 \emph{Phys.
Rev.} A \textbf{89} 063816

\bibitem{nori-n} Wilson C M \emph{et al.} 2011 \emph{Nature} \textbf{479} 376

\bibitem{me-arx} Dodonov A V \emph{et al.} 2008 arXiv:0806.4035v3

\bibitem{jpcs} Dodonov A V 2009 \emph{J. Phys.: Conf. Ser.} \textbf{161}
012029

\bibitem{liberato9} De Liberato S \emph{et al.} 2009 \emph{Phys. Rev.} A
\textbf{80} 053810

\bibitem{2level} Dodonov A V and Dodonov V V 2012 \emph{Phys. Rev.} A
\textbf{85} 015805

\bibitem{2atom} Dodonov A V and Dodonov V V 2012 \emph{Phys. Rev.} A \textbf{%
85} 055805

\bibitem{3level} Dodonov A V and Dodonov V V 2012 \emph{Phys. Rev.} A
\textbf{85} 063804
\end{thebibliography}
\end{document}